\documentclass[preprint,12pt]{elsarticle}

\usepackage{amsthm, amsmath, amssymb, mathrsfs, bm}
\usepackage{tikz}
\usepackage{xcolor}

\usepackage{graphicx, float, subcaption, diagbox, booktabs, tabularx, array, multirow, threeparttable, longtable}

\usepackage[utf8]{inputenc}
\usepackage{newtxtext, newtxmath}
\usepackage[english]{babel}
\usepackage{microtype}
\usepackage[section]{placeins}
\usepackage{placeins}

\usepackage{geometry}
\usepackage{algorithm}
\usepackage{algorithmic}

\usepackage{siunitx}

\biboptions{sort&compress}  % 连号压缩成 [31-34]
\usepackage[unicode, colorlinks=true, linkcolor=blue, citecolor=green, urlcolor=blue]{hyperref}

\graphicspath{{figures/}}

\makeatletter
\newcommand\dif{%
  \mathop{}\!%
  \ifthu@math@style@TeX
    d%
  \else
    \mathrm{d}%
  \fi
}
\makeatother

\journal{Elsevier}

\begin{document}
\begin{frontmatter}

% \title{An ALE-informed Machine Learning Framework for Predicting Fluid-Structure Interaction}
\title{An ALE-Consistent Graph Neural Operator-Transformer Framework for Fluid-Structure Interaction}

% \author{Shihang Zhaog (赵世航)\textsuperscript{a}, Martín Saravia\textsuperscript{b}, Zhiyang Xue (薛智洋)\textsuperscript{a}, Haokui Jiang (江昊葵)\textsuperscript{a}, and Shunxiang Cao (曹顺翔)\textsuperscript{a,*}}

\author{Shihang Zhao\textsuperscript{a}, Martín Saravia\textsuperscript{b}, Haokui Jiang\textsuperscript{a}, Zhiyang Xue\textsuperscript{a}, and Shunxiang Cao\textsuperscript{a,*}}

\affiliation{organization={Institute for Ocean Engineering},
            addressline={Shenzhen International Graduate School, Tsinghua University}, 
            city={Shenzhen},
            state={Guangdong},
            postcode={518055}, 
            country={China}}
\affiliation{organization={Centro de Investigación en Mecánica Teórica y Aplicada, Universidad Tecnológica Nacional - CONICET},
            addressline={11 de Abril 461}, 
            city={Bahía Blanca},
            country={Argentina}}

% 我们提出一种用于流固耦合（FSI）预测的、\textbf{满足任意拉格朗日-欧拉（ALE）一致性}的图神经算子—Transformer 框架。该代理模型由三部分组成：用于非结构网格的图神经算子、用于推进流体时空预测的视觉 Transformer，以及用于预测结构边界运动学量的轻量级长短期记忆网络（LSTM）。各模型通过标准的 ALE 数据交换流程进行耦合。为在移动界面处强化运动学相容性，本文引入一种 \textbf{ALE 一致性的边界修正（boundary correction）}：在每次耦合更新时，用结构侧预测的界面速度覆写流体侧界面速度。训练方面采用两阶段策略——先进行单步预测预训练，再进行长时序自回归微调——以抑制滚动预测中的误差累积。
% 在 Turek-Hron 圆柱-梁基准算例上，该框架能够准确预测周期与非周期两类 FSI 响应，捕捉变形边界及其沿界面的压力诱导载荷演化。同时，该方法在入口型线变化的插值与外推测试中均表现出良好的泛化能力。
% 在长时序滚动推演中，该框架可保持接近 1 的决定系数；而移除边界修正会使 $R^2$快速下降至 0.85，表明 ALE 一致性的运动学约束对维持长期预测精度与稳定性至关重要。结果表明 \textbf{ALE 一致性耦合}与\textbf{长时序训练}是构建稳定、准确且具可迁移性的 FSI 代理模型的关键，为实时控制与快速设计研究提供了可行基础。
            
\begin{abstract}  
We propose an arbitrary Lagrangian-Eulerian (ALE)-consistent machine learning framework for long-term fluid-structure interaction (FSI) prediction on deforming unstructured meshes. Specifically, the fluid dynamics are modeled by a surrogate that combines a graph neural operator (GNO) with a vision Transformer (ViT) for spatiotemporal prediction, while a lightweight long short-term memory (LSTM) network predicts structural kinematics at the interface. The two surrogates are coupled through a standard partitioned procedure. Most importantly, kinematic compatibility at the moving interface is enforced via an ALE-consistent boundary-correction step that updates the fluid-side interface velocity with the predicted structural velocity at each coupling update, thereby improving near-interface accuracy and long-term rollout stability.
To mitigate autoregressive error accumulation,
a two-stage training strategy is adopted, consisting of single-step supervised pretraining followed by long-term autoregressive fine-tuning.
The proposed framework is validated on the benchmark problem of a flexible beam vibration in the wake of a cylinder. Results demonstrate accurate phase-consistent predictions over long rollouts and robust generalization under inlet-profile variations in both interpolation and extrapolation settings. Systematic ablation studies further assess the respective contributions of the ViT module, ALE-consistent boundary correction, and long-term training to predictive accuracy and rollout robustness. 

% The proposed framework is validated on a flexible beam vibration in the wake of a cylinder and accurately predicts both periodic and non-periodic FSI responses, capturing interface deformation and pressure-induced loading. Moreover, it generalizes robustly across inlet-profile variations under both interpolation and extrapolation. 

% Across rollouts, the framework maintains an $R^2$ close to 1, whereas removing the boundary-correction rapidly reduces $R^2$ to 0.85, highlighting the necessity of ALE-consistent kinematic enforcement. Likewise, long-term fine-tuning improves rollout stability, increasing the end-of-term $R^2$ from 0.95 to 0.99.

\end{abstract}


\end{frontmatter}

\section{Introduction}
\label{sec:Intro}

%%%%%%%%%%
%流固耦合应用广泛，计算方法主要有IBM和ALE，ALE更加贴合工程实际；计算速度慢，特别是双向流固耦合，引出数据驱动方法可以加速此过程；
%传统降阶和机器学习对比，机器学习非线性处理能力强，此处综述灰盒
%综述机器学习流固耦合的研究，分为非定常流动和动边界处理两部分，指出长时预测累积误差和柔性边界变形问题
%
%本文受ALE计算方法启发，搭建了GNO-Vit预测流体，LSTM预测固体的流固耦合计算框架，GNO处理柔性边界变形，边界修正提高预测精度，长时序列训练提升长时预测能力。
%%%%%%%%%%

% 流固耦合（Fluid-Structure Interaction, FSI）广泛存在于海洋工程与航空航天等工程场景中~\cite{huera2025vortex,kou2021data}，流固耦合的快速可靠预测是安全评估与性能优化的关键。工程实践中，FSI 的高保真数值模拟主要依托浸入边界法（Immersed Boundary Method, IBM）与任意拉格朗日-欧拉法（Arbitrary Lagrangian-Eulerian, ALE）两类框架~\cite{peskin2002immersed,verzicco2023immersed,souli2013arbitrary,组内文章}。其中，ALE 采用随结构运动的贴体变形网格，能够更严格地施加界面运动学条件并降低边界截断误差，从而通常具备更高的近壁预测精度，因而在工业界得到广泛应用~\cite{hou2012numerical,qian2024overview,组内文章}。然而，高精度的双向耦合往往以较高的计算代价为代价：在分区求解策略下，每一时间步内需要在流体与结构子问题之间进行界面数据交换，并通过多次子迭代满足界面相容性与收敛要求~\cite{morab2020overviewcomputationalfluidstructure,组内文章}。该迭代耦合过程相较单一物理场计算显著更为昂贵，且在大变形、强非线性工况下迭代次数进一步增加，导致整体计算成本迅速攀升，从而制约了需要大量重复求解的高阶任务，如参数优化~\cite{haubner2024numerical}、实时控制~\cite{nair2023bio,组内文章}与不确定性量化~\cite{beran2017uncertainty}等。基于此，亟需发展在保持工程所需精度的同时显著降低计算成本的 FSI 代理模型；而近年来数据驱动建模的进展为加速流固耦合预测提供了可行路径~\cite{afridi2024fluid,brunton2020machine,herrmann2024deep}。

Fluid-structure interaction (FSI) plays a central role in a wide range of engineering applications, particularly in ocean, aerospace, and biomedical engineering~\cite{huera2025vortex,wang2023numerical,kou2021data,sun2025modeling}. Some emerging applications, such as digital twin~\cite{cao2022bayesian}, uncertainty quantification~\cite{beran2017uncertainty},and real-time active control for FSI~\cite{nair2023bio,liu2024novel_r,zhao2025comparative,jiang2026model,yin2026co}, increasingly demand fast and repeatable FSI simulations. However, conventional high-fidelity FSI simulations are fundamentally ill-suited for such time-critical scenarios due to their high computational cost. This limitation lies in the multi-physics nature of FSI problems, which requires both tight temporal coupling between fluid and structural solvers as well as accurate spatial treatment of deforming fluid–structure interfaces. Specifically, the partitioned procedure is widely used for two-way FSI coupling, in which the fluid and structural subproblems are solved separately and synchronized through interfacial data exchange at each time step~\cite{morab2020overviewcomputationalfluidstructure}. However, achieving stable and convergent solutions generally requires multiple sub-iterations, with the computational overhead increasing markedly in the presence of strong added-mass effects~\cite{causin2005added,cao2018robin}. As for spatial coupling, conventional FSI simulations commonly rely on
arbitrary Lagrangian-Eulerian (ALE) method~\cite{souli2013arbitrary} to represent the motion and deformation of fluid–structure interfaces. By employing body-fitted, deforming meshes that moves with the structure, ALE enables accurate enforcement of interface conditions and superior near-wall accuracy. However, the need to continuously update body-fitted moving meshes substantially increases
computational complexity. These computational bottlenecks motivate the development of FSI surrogate models that retain engineering-level accuracy while substantially reducing computational expense. 

Recent advances in data-driven modeling offer a promising direction toward accelerating FSI simulation~\cite{afridi2024fluid,brunton2020machine,herrmann2024deep}. 
% 现有的数据驱动 FSI 预测方法总体可归纳为两条技术路线：传统降阶模型（reduced-order models, ROMs）~\cite{taira2017modal}以及机器学习方法~\cite{brunton2020machine}。前者通常从高保真数值模拟或实验数据中提取主导空间模态，并通过投影将高维流场/结构场演化压缩为低维动力系统，从而实现计算加速~\cite{taira2017modal，组内文章}。典型代表包括基于 Galerkin 投影的 POD 模型（Galerkin-POD）~\cite{whisenant2020galerkin}、面向输入-输出系统的平衡截断/平衡 POD（BPOD）~\cite{jiang2024balanced}，以及基于 Koopman 算子谱分解的线性表示方法~\cite{jiang2025koopman}，它们在周期性或近线性响应、以及小扰动范围内往往能够以较低的维数重构主要动力学特征。然而，工程 FSI 常伴随界面大变形、涡脱落与再附着、非定常载荷突变等强非线性过程，使得低阶子空间可能随工况显著变化，进而导致固定基底的 ROM 难以兼顾精度与稳健性；同时，ROM 的建模与闭合往往需要显式引入控制方程、投影形式与经验闭合项，模型结构对特定算例与参数区间具有较强依赖性，从而限制了其跨工况迁移与泛化能力。
% 相比之下，ML 方法（尤其是深度学习）能够直接从数据中学习高维状态到未来状态的非线性映射，具有端到端特征提取与高效推理的优势~\cite{brunton2020machine}。一旦训练完成，模型可在极低计算成本下进行快速预测，因而在复杂流动演化建模、结构响应估计~\cite{kochkov2021machine,kou2021data} 以及流固耦合代理建模~\cite{ong2025synergizing,tiba2025machine} 等方面表现出显著潜力。
Existing studies have primarily explored two methodological directions: (i) reduced-order models (ROMs) that leverage physics-based modeling and dimensionality reduction~\cite{taira2017modal,sharma2024lagrangian,prakash2024projection}, and (ii) end-to-end machine learning (ML) frameworks that directly learn predictive mappings from data~\cite{brunton2020machine}. ROMs typically extract dominant spatial modes from experiments or high-fidelity simulations and project the high-dimensional fluid/structure evolution onto a low-dimensional dynamic system. This reduction enables substantial computational acceleration while retaining physical interpretability~\cite{taira2017modal}. Representative examples include Galerkin projected POD models~\cite{whisenant2020galerkin}, balanced truncation POD for input-output systems~\cite{jiang2024balanced}, and Koopman operator-based spectral representations that seek linear embedding of nonlinear dynamics~\cite{jiang2026model, jiang2025koopman}. Such ROMs often perform well for periodic or weakly nonlinear responses and within small perturbation regimes, where the dominant dynamics can be captured by a compact subspace. However, practical FSI problems frequently involves strong nonlinearities, including large structural deformation, vortex shedding, and abrupt variations of unsteady loads. Under these conditions, the relevant low-dimensional subspace may vary significantly with operating parameters, leading to a trade-off between accuracy and robustness. Moreover, ROMs construction and closure typically rely on explicit use of governing equations, projection formulations, and empirical closure terms~\cite{taira2017modal,brunton2020machine}. As a result, their performance can be highly sensitive to the specific configuration and parameter range, limiting their transferability and generalization across different flow regimes.

In contrast, ML approaches, particularly deep learning, aim to learn nonlinear mappings from high-dimensional system states to future responses in an end-to-end manner, leveraging automatic feature extraction and fast inference~\cite{brunton2020machine}. Representative models include encoder–decoder and sequence-learning architectures built on convolutional neural networks (CNNs), recurrent units, or attention mechanisms~\cite{liu2024novel,fan2024differentiable,rahman2024pretraining}. Once trained, such models can provide rapid predictions at very low computational cost, which has driven their success in complex flow predicting~\cite{barwey2025mesh,sousa2024enhancing}, structural response estimation~\cite{kochkov2021machine,kou2021data}, and FSI surrogate modeling~\cite{ong2025synergizing,tiba2025machine}.

% 但是，基于机器学习的流固耦合（FSI）预测在工程可用性上仍面临两项核心瓶颈：一是长时序自回归预测中的误差累积与相位漂移，这是非定常流动与耦合系统预测中普遍存在且最易导致滚动预测失稳的问题~\cite{han2019novel}；二是结构边界的变形与运动所引入的几何非定常性与界面约束，使得模型不仅要学习流场与结构场各自的演化规律，还必须在界面处持续满足运动学相容与动力学平衡，从而显著提高了学习与泛化难度~\cite{cheng2025machine}。因为这些问题的存在，误差累积会在非周期/强瞬态工况中被放大，因此现有多数工作仍主要集中于周期性响应或准周期振荡~\cite{raissi2019deep, gao2024predicting, zhang2022data, tang2022transfer, yin2024physics}；针对非周期动力学的研究相对有限，且在强瞬态阶段的预测精度与稳定性仍不足~\cite{fan2024differentiable, zhai2025projection}。

However, ML-based FSI surrogates still face two fundamental bottlenecks that limit their applicability in engineering problems. The first is the accumulation of autoregressive errors and the associated phase drift in long-term rollouts, a pervasive issue in unsteady flow prediction and coupled dynamical systems that often leads to instability over time~\cite{han2019novel}. The second arises from the deforming fluid-structure interface, which leads to geometric non-stationarity and strict interfacial constraints. Beyond learning the separate evolutions of the fluid and structural states themselves, the surrogate must continuously satisfy kinematic compatibility and dynamic equilibrium at the moving interface, substantially increasing the difficulty of learning and generalization~\cite{cheng2025machine}. These challenges become particularly severe in non-periodic and strongly transient regimes, where error accumulation is amplified. Consequently, most existing studies still focus primarily on periodic or quasi-periodic responses~\cite{raissi2019deep,gao2024predicting,zhang2022data,tang2022transfer,yin2024physics}, while attempts to predict non-periodic dynamics remain limited and the accuracy and stability in strongly transient phases are often insufficient~\cite{fan2024differentiable,zhai2025projection}.

% 具体的，对于第一项挑战，已有研究提出了多种缓解思路，包括将有限差分/有限体积等数值离散格式嵌入网络以注入物理先验~\cite{fan2024differentiable}、采用一次性多步预测以减少预测次数~\cite{du2024conditional, zhou2024text2pde}，以及通过长时序训练（long-term training）在训练阶段显式“暴露”模型于其自身误差传播过程~\cite{fan2024differentiable}。其中，离散格式嵌入能够在一定程度上约束时间推进并降低误差扩散，但对数据噪声与数值一致性较为敏感。多步预测策略在纯流动问题或边界运动预先给定情形下较为有效，但对于 ALE 类需要持续双向耦合的 FSI，界面运动由预测载荷驱动且必须逐步更新，直接采用固定多步预测往往难以保持界面一致性与耦合闭环的稳定性。相比之下，长时序训练通过在有限滚动窗口内最小化多步累计误差，能够更贴合 FSI 的闭环自回归推演特点：模型不仅优化单步局部误差，更直接学习抑制误差在耦合反馈链中的放大机制。
% 然而，从零开始进行长时序训练通常会带来显著的优化困难：初始模型误差较大时，滚动输入迅速偏离真实数据分布，导致梯度信号不稳定并引发收敛失败或训练塌陷。针对这一问题，本文采用与工程训练实践更为匹配的两阶段策略：首先以单步监督进行预训练，获得具备可靠局部预测能力的初始化；随后在此基础上开展长时序自回归微调，使模型在保持单步精度的同时学习抵抗误差累积与相位漂移，从而显著提升收敛稳定性与长时序预测性能。

To mitigate autoregressive error accumulation, several strategies have been explored. One class of approaches seeks to regularize temporal evolution by embedding numerical discretization operators, such as finite-difference or finite-volume schemes, into neural networks to incorporate physical priors~\cite{fan2024differentiable}. While such methods can suppress error growth, they are often sensitive to data noise and discretization consistency. An alternative strategy is to predict multiple time steps per inference to reduce the number of recursive evaluations~\cite{du2024conditional,zhou2024text2pde}. This approach has proven effective in predicting unsteady flow dynamics with prescribed structural motion. However, in two-way coupled FSI, where interface motion must be updated sequentially in response to evolving fluid loads, the use of fixed multi-step prediction may lead to interface inconsistencies and reduced coupling stability. A more promising approach is to adopt long-term training in which the model is explicitly exposed to its own error propagation during optimization~\cite{fan2024differentiable}. By minimizing accumulated multi-step errors over a finite rollout window, this approach directly learns to suppress error amplification along the coupling feedback chain, rather than optimizing only one-step local errors. Therefore, it is better aligned with the closed-loop, autoregressive nature of FSI. Nevertheless, optimizing long-term objectives from scratch remains challenging. When the initial model is insufficiently accurate, autoregressive inputs can rapidly drift away from the training distribution, leading to unstable gradients and poor convergence or training collapse. To address this challenge, we adopt a two-stage training protocol that is more consistent with engineering practice. Specifically, single-step supervised pretraining is first employed to obtain a reliable initialization with strong local accuracy. The model is then fine-tuned using long-term autoregressive training, enabling it to resist error accumulation and phase drift while preserving one-step fidelity. This staged strategy significantly improves convergence stability and long-term predictive performance.

% 针对第二个问题，现有研究主要沿四条路线展开：将 IBM 与学习模型结合以避免网格变形~\cite{fan2024differentiable,xiao2024fourier}；将非结构数据投影/插值到结构化网格或统一表示空间以接入图像式网络~\cite{liu2024novel,gupta2022hybrid}；通过网格展开（mesh unfolding）等几何映射保留欧拉网格结构~\cite{han2022deep}；以及直接在原始离散节点/单元上构建图结构，采用图方法处理非结构数据~\cite{rahman2024pretraining,horie2022physics,pfaff2020learning,gao2024finite}。其中，IBM+ML 虽能规避网格变形，但界面信息往往以不规则形式注入网络，仍需要复杂的边界处理与插值/传播算子，容易引入近壁偏差并影响载荷传递稳定性。投影/插值方法可以实现与图像框架的无缝对接，但其核心难点转化为界面双向传递的保真性：若力与位移（或速度）在界面上的映射稍有误差，便会直接污染近壁压力/剪切并通过耦合反馈放大误差。网格展开方法对规则构型有效，但面对工程中常见的非结构或自适应网格时可扩展性受限~\cite{liu2024novel}。相比之下，图方法可在原始网格上自然表达几何与拓扑关系，尤其图神经算子（GNO）在复杂几何与非结构网格的流场建模中显示出优势~\cite{cheng2025machine,anandkumar2020neural,li2023geometry}。
% 然而，现有多数图方法仍主要聚焦于短时预测或预设结构运动条件下的流场重构~\cite{gao2024predicting}，缺乏对真实双向耦合机制的显式建模。

To address the second challenge associated with deforming interfaces, existing studies have primarily explored four directions. A straightforward method is to combine learning models with immersed boundary methods to avoid mesh deformation~\cite{fan2024differentiable,xiao2024fourier}. However, the incorporation of interface information still relies on nonconforming boundary treatments and interpolation, which can introduce near-wall bias and degrade the accuracy of load transfer across the interface. An alternative line of work seeks to project unstructured data onto structured grids or unified latent representations~\cite{liu2024novel,gupta2022hybrid}. This strategy enables the use of image-based networks, but it suffers from robustness issue of interfacial coupling. Even small mapping errors in forces, displacements and velocities at the interface can directly contaminate near-wall pressure and be amplified through the coupled feedback. Another class of methods preserves an Eulerian grid structure through geometric mappings such as mesh unfolding~\cite{han2022deep}. While effective for relatively regular geometries, their scalability is limited for the unstructured or adaptive meshes commonly encountered in engineering applications~\cite{liu2024novel}. In contrast to the approaches described above, a more principled solution is to construct graph representations directly on the original discretization and perform graph-based learning on unstructured meshes~\cite{rahman2024pretraining,horie2022physics,pfaff2020learning,gao2024finite}. By operating on the original nodes, these methods naturally encode geometric and topological information without requiring mesh regularization or geometric remapping. In particular, graph neural operators (GNOs) have demonstrated clear advantages for flow modeling under such settings~\cite{cheng2025machine,anandkumar2020neural,li2023geometry}. However, most existing graph-based approaches still focus on short-term prediction or flow evolution under prescribed structural motion~\cite{gao2024predicting}, and are therefore not directly applicable to fully two-way coupled FSI problems.

% 为应对以上问题，本文提出一种用于变形网格上耦合流固动力学预测的 {ALE-consistent} 图神经算子-Transformer 框架。具体而言，流体演化由一种混合代理模型刻画：在非结构网格上采用图神经算子（GNO）进行空间建模，并结合基于 Transformer 的时空预测器以高效推进流场状态。与此同时，结构边界的运动学量由轻量级长短期记忆网络（LSTM）进行预测，利用结构响应维度相对较低的特点实现高效推理。流体与结构预测器通过标准的 ALE 数据交换流程进行耦合，并引入 ALE 一致性的边界修正步骤：在每次耦合更新时，使用结构预测得到的界面速度覆写流体侧界面速度，从而在运动界面处强制满足运动学相容性。为提升长时序推演的稳定性，本文采用两阶段训练策略对代理模型进行训练：先进行单步预测预训练，再进行长时序自回归微调，以缓解滚动预测中的误差累积与相位漂移。
% 本文在 Turek-Hron 圆柱-梁基准算例上验证了所提框架的有效性，评估了周期与非周期两类 FSI 工况，并通过入口型线变化的插值与外推测试进一步考察其泛化能力。此外，本文还开展消融实验，定量分析 ALE 一致性边界修正与长时序训练在稳定长期预测与提升近壁预测保真度方面的作用。
% 本文结构安排如下：第~\ref{sec:Framework}~节给出控制方程、ALE 耦合数据交换与所提出的 ALE-consistent 学习框架；第~\ref{sec:Results}~节展示在基准算例上的预测结果与消融分析，并进一步验证非周期工况下的长期稳定性与泛化能力；第~\ref{sec:Conclusions}~节总结本文主要结论并讨论后续工作方向。

Motivated by these limitations, we develop an ALE-consistent graph neural operator-Transformer framework for fast prediction of FSI problems involving large deformation of flexible structures. Specifically, the fluid dynamics is modeled by a hybrid surrogate that combines a GNO defined on unstructured meshes for spatial modeling with a vision Transformer (ViT)-based spatiotemporal predictor. Meanwhile, the structural boundary kinematics are predicted using a lightweight long short-term memory (LSTM) network, leveraging the relatively low-dimensional nature of the structural response to enable efficient inference. The fluid and structural predictors are coupled through the standard ALE-based partitioned procedure, augmented by an ALE-consistent boundary correction: at each coupling update, the fluid-side interface velocity is corrected by the predicted structural interface velocity, thereby enforcing kinematic compatibility at the moving interface. To mitigates long-term error accumulation and phase drift, a two-stage training protocol is employed, which consists of single-step pretraining followed by long-term autoregressive fine-tuning. 

The proposed framework is validated on the benchmark problem of a flexible beam vibration in the wake of a cylinder~\cite{turek2006proposal}. Both periodic and non-periodic FSI regimes are considered, and the generalization performance is further assessed under inlet-profile variations through interpolation and extrapolation tests. In addition, ablation studies are conducted to quantify the respective contributions of the ViT-based temporal modeling, the ALE-consistent boundary correction, and long-term autoregressive fine-tuning to rollout stability and near-boundary prediction fidelity.
The remainder of the paper is organized as follows. Section~\ref{sec:Framework} presents the governing equations and the proposed framework. Section~\ref{sec:Results} reports the numerical results and ablation analyses, including the predictions in the non-periodic regime. Section~\ref{sec:Conclusions} summarizes the main conclusions.

%%%%%%%%%%%%%%%%%%%%%%%%%%%%%%%%%%%%%%%%%%%%%%%%%%%%%%%%%%%%%%%%%%%%%%%%%

\section{Methodology}
\label{sec:Framework}
This section describes the physical model and the proposed GNO-ViT framework. We first present the ALE-based governing equations for FSI, which are used to generate the datasets and define the interface constraints. The hybrid surrogate architecture and its ALE-consistent coupling workflow are then described, including the boundary-correction step that enforces interfacial kinematic compatibility. Finally, the two-stage training protocol adopted to stabilize rollouts and mitigate error accumulation is detailed.

\subsection{Problem formulation}
\label{subsec:FOM}

\begin{figure}
    \centering
    \includegraphics[width=0.75\linewidth]{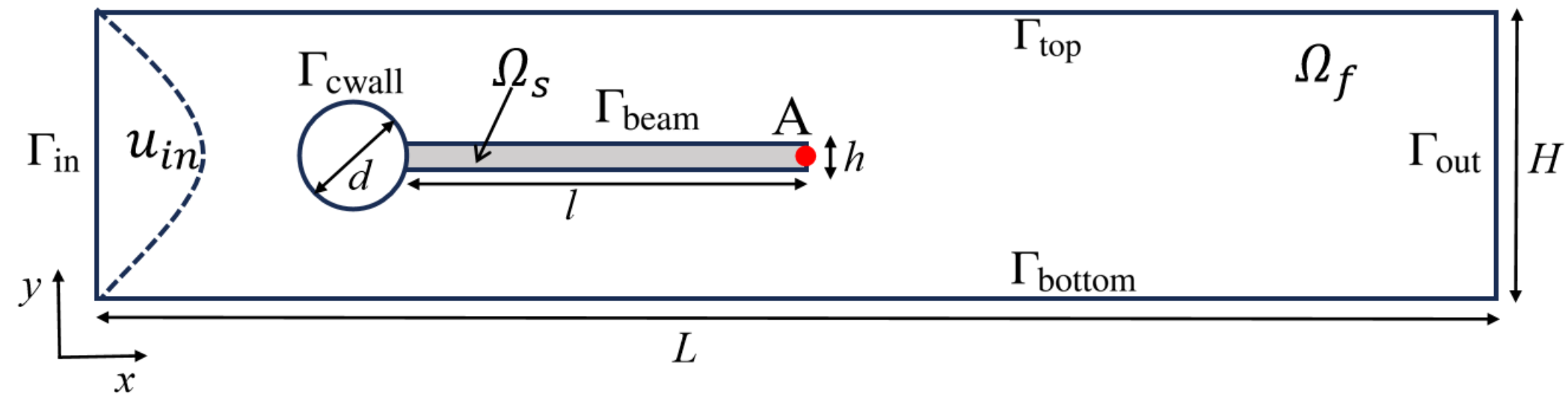}
    \caption{Schematic of the flexible beam (gray shaded region) clamped to a rigid cylinder (white circle) and immersed in a channel flow.}
    \label{fig:Fig-case}
\end{figure}

We consider the classical FSI benchmark of a flexible beam vibration in the wake of a cylinder~\cite{turek2006proposal} (Figure~\ref{fig:Fig-case}). 
The FSI system consists of a fluid subproblem, a structure subproblem, a mesh-update problem, and interface conditions~\cite{jiang2026model}. Specifically, the fluid dynamics is governed by the incompressible Navier-Stokes equations, which are formulated in an arbitrary Lagrangian-Eulerian framework.
\begin{equation}
\nabla\cdot\bigl[\boldsymbol{\Phi}(\boldsymbol{\eta}_e)\,\mathbf{u}_f\bigr]=0,
\qquad \text{in } \Omega_f,
\label{eq:NS1}
\end{equation}
\begin{equation}
\rho_f\,J(\boldsymbol{\eta}_e)\,\frac{\partial \mathbf{u}_f}{\partial t}
+
\rho_f\,\bigl[(\nabla \mathbf{u}_f)\,\boldsymbol{\Phi}(\boldsymbol{\eta}_e)\bigr]
\left(\mathbf{u}_f-\frac{\partial \boldsymbol{\eta}_e}{\partial t}\right)
-
\nabla\cdot \boldsymbol{\Sigma}(\mathbf{u}_f,p_f,\boldsymbol{\eta}_e)
=0,
\qquad \text{in } \Omega_f,
\label{eq:NS2}
\end{equation}
where $\boldsymbol{\eta}_e$ is the displacement field describing the deformation of the fluid mesh, $\mathbf{u}_f,\rho_f,p_f$ are the fluid velocity, density, and pressure, respectively. 
$\boldsymbol{\Phi}(\boldsymbol{\eta}_e)=J(\boldsymbol{\eta}_e)\bigl(\mathbf{I}+\nabla \boldsymbol{\eta}_e\bigr)^{-1}$ is the ALE deformation operator and $J(\boldsymbol{\eta}_e)=\det\!\bigl(\mathbf{I}+\nabla \boldsymbol{\eta}_e\bigr)$ is the Jacobian of the mapping. The first Piola-Kirchhoff stress tensor for the fluid is given by
\begin{equation}
\boldsymbol{\Sigma}(\mathbf{u}_f,p_f,\boldsymbol{\eta}_e)
=
\left\{
-p_f\,\mathbf{I}
+
\frac{\mu_f}{J(\boldsymbol{\eta}_e)}
\left[
(\nabla\mathbf{u}_f)\,\boldsymbol{\Phi}(\boldsymbol{\eta}_e)
+
(\nabla\mathbf{u}_f)^{\mathsf{T}}\boldsymbol{\Phi}(\boldsymbol{\eta}_e)^{\mathsf{T}}
\right]
\right\}\boldsymbol{\Phi}(\boldsymbol{\eta}_e)^{\mathsf{T}},
\label{eq:fluid_PK1}
\end{equation}
where $\mu_f$ is the fluid dynamic viscosity.

The structural dynamics is modeled by a compressible Saint-Venant-Kirchhoff (SVK) model described in the Lagrangian framework, which reads:
\begin{equation}
\rho_s\,\frac{\partial^2 \boldsymbol{\eta}_s}{\partial t^2}
-
\nabla\cdot\Bigl[\mathbf{F}(\boldsymbol{\eta}_s)\,\mathbf{S}(\boldsymbol{\eta}_s)\Bigr]
=0,
\qquad \text{in } \Omega_s,
\label{eq:structure}
\end{equation}
where $\boldsymbol{\eta}_s$ denote the structure displacement field, $\rho_s$ is the structural density, and
$\mathbf{F}(\boldsymbol{\eta}_s)=\mathbf{I}+\nabla\boldsymbol{\eta}_s$ is the deformation gradient. and $\mathbf{S}(\boldsymbol{\eta}_s)$ is the second Piola-Kirchhoff stress tensor given by
\begin{equation}
\mathbf{S}(\boldsymbol{\eta}_s)
=
\lambda_s\,\mathrm{tr}\!\bigl(\mathbf{E}(\boldsymbol{\eta}_s)\bigr)\mathbf{I}
+
2\mu_s\,\mathbf{E}(\boldsymbol{\eta}_s).
\label{eq:structure_SVK}
\end{equation}
where $\lambda_s$ and $\mu_s$ are the Lam\'e parameters, 
$\mathbf{E}(\boldsymbol{\eta}_s)=\frac{1}{2}\bigl[\mathbf{F}(\boldsymbol{\eta}_s)^{\mathsf{T}}\mathbf{F}(\boldsymbol{\eta}_s)-\mathbf{I}\bigr]$ is the Green-Lagrange strain.

To smoothly extend the interface displacement into the interior fluid mesh under potentially large deformations, a standard ALE approach solves the mesh-extension problem with interfacial continuity:
\begin{equation}
\nabla\cdot \boldsymbol{\Sigma}_e(\boldsymbol{\eta}_e)=0,
\qquad \text{in } \Omega_e,
\label{eq:extension}
\end{equation}
\begin{equation}
\boldsymbol{\eta}_e=\boldsymbol{\eta}_s,
\qquad \text{on } \Gamma_{\text{beam}},
\label{eq:extension_bc}
\end{equation}
where $\boldsymbol{\Sigma}_e$ denotes the chosen extension operator and $\Omega_e$ is the interior of the deformable fluid-mesh domain, and $\Gamma_{\text{beam}}$ denotes the wetted surface of the beam. 
In our implementation, the fluid mesh is updated via Gmsh remeshing driven by the updated interface boundary points, so the continuity in Eq.~\eqref{eq:extension_bc} is satisfied by construction.

The fluid and structure domains are coupled on the interface through kinematic and dynamic conditions:
\begin{equation}
\mathbf{u}_f-\frac{\partial \boldsymbol{\eta}_e}{\partial t}=0,\quad
\text{on } \Gamma_{\text{beam}},
\label{eq:kinematic_IC}
\end{equation}
\begin{equation}
\boldsymbol{\Sigma}(\mathbf{u}_f,p_f,\boldsymbol{\eta}_e)\,\mathbf{n}
-
\Bigl[\mathbf{F}(\boldsymbol{\eta}_s)\,\mathbf{S}(\boldsymbol{\eta}_s)\Bigr]\mathbf{n}
=0,
\quad
\text{on } \Gamma_{\text{beam}},
\label{eq:dynamics_IC}
\end{equation}
where $\mathbf{n}$ denotes the outward unit normal vector on the structural surface. The first condition enforces velocity continuity at the interface and the second ensures traction equilibrium. Eqs.~\eqref{eq:extension}, ~\eqref{eq:extension_bc},~\eqref{eq:kinematic_IC} and~\eqref{eq:dynamics_IC} provide the physical basis for the ALE-consistent boundary correction used in our developed framework.

To interface with the data representation used in the developed framework, we introduce the following shorthand notation for fluid mesh deformation and on the FSI interface $\Gamma_{\text{beam}}$:
\begin{equation}
\mathbf{d}_f := \boldsymbol{\eta}_e,\qquad
\mathbf{u}_f = \frac{\partial \mathbf{d}_f}{\partial t},\qquad
\mathbf{d}_s := \boldsymbol{\eta}_s\big|_{\Gamma_{\text{beam}}},\qquad
\mathbf{u}_s =\frac{\partial \mathbf{d}_s}{\partial t},
\label{eq:notation_bridge}
\end{equation}
where $(\mathbf{d}_f,\mathbf{u}_f)$ denotes the mesh displacement and velocity field in the ALE formulation, while $(\mathbf{d}_s,\mathbf{u}_s)$ denote the interfacial structural displacement and velocity, respectively. Unless otherwise stated, the remainder of this paper adopts the shorthand in Eq.~\eqref{eq:notation_bridge}. This is a notational change only and does not modify the governing equations or interface constraints.

\subsection{Computational framework and ALE-consistent coupling scheme}
\label{subsec:framework}
\begin{figure}
    \centering
    \includegraphics[width=1.0\linewidth]{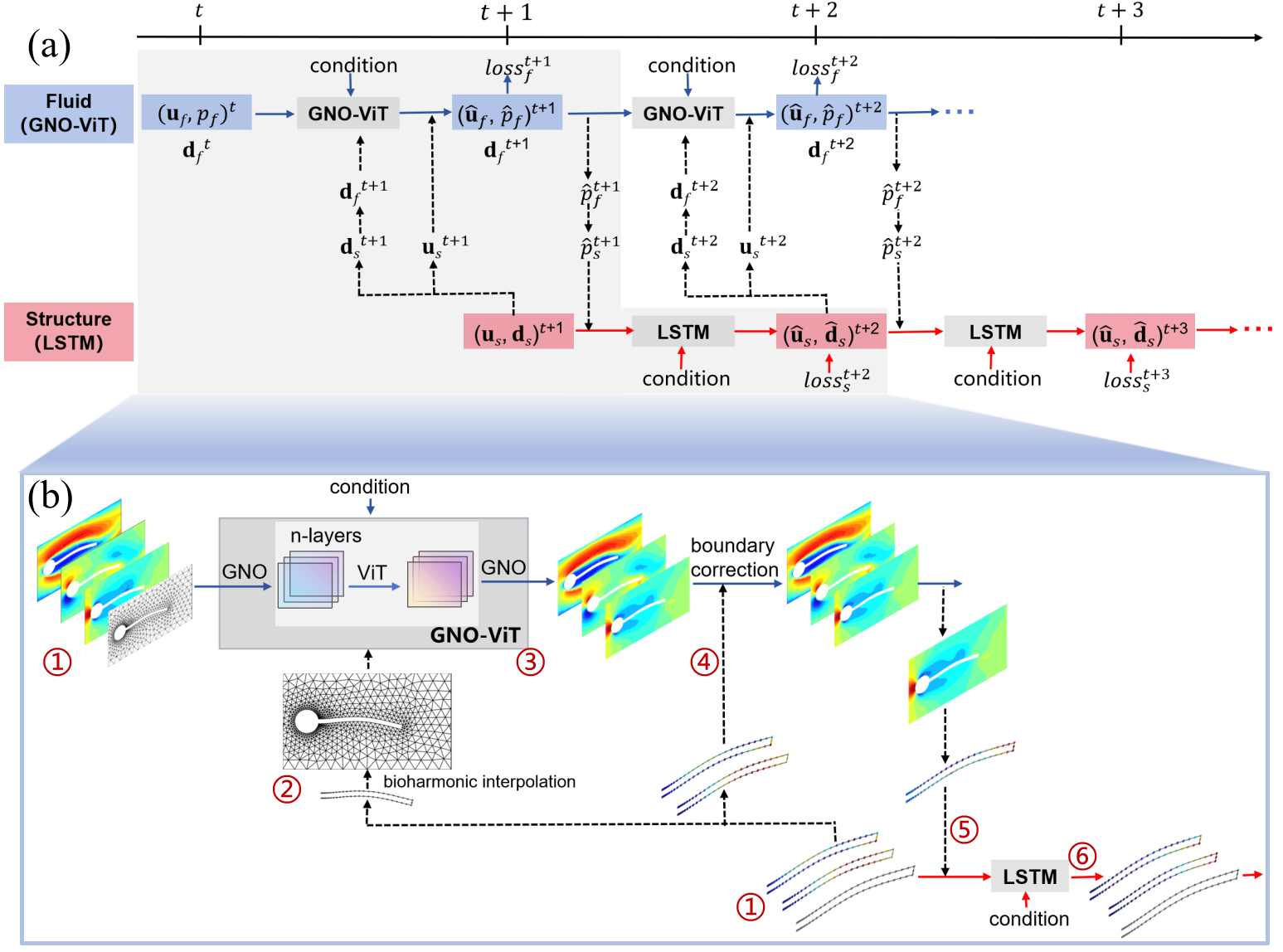}
    \caption{
    ALE-consistent GNO--Transformer framework for long-horizon FSI prediction.
    (a) Overview of the autoregressive rollout, where the fluid surrogate (GNO--ViT) and the structure surrogate (LSTM) are advanced in a staggered manner and supervised by step-wise losses to suppress error accumulation.
    (b) Detailed schematic of one rollout cycle with red circled indices (1-6) corresponding to the six major steps in Algorithm~\ref{alg:computation-process}:
     Initialization, Mesh update, Fluid prediction, ALE-consistent boundary correction, Pressure transfer to structure and Structure prediction.
    Dashed arrows indicate the coupling information exchanged between the two surrogates, and $\mathrm{loss}_f^{t}$ and $\mathrm{loss}_s^{t}$ denote the step-wise training losses used in the two-stage training protocol.
    }
    \label{fig:vit-gno}
\end{figure}

The proposed framework follows the data-exchange procedure of a conventional partitioned ALE-based FSI solver, while the computationally intensive fluid and structural sub-problems are replaced by trained surrogate models. For brevity, we refer to the proposed framework as GNO-ViT in the remainder of this paper. The overall ALE-consistent time-marching algorithm of the framework is illustrated in Figure~\ref{fig:vit-gno} and summarized in Algorithm~\ref{alg:computation-process}.

For ease of description, we define the fluid state on the deforming mesh as
\begin{equation}
F^{t}=\bigl(\mathbf{q}_f, \mathbf{d}_f\bigr)^{t},
\quad \text{where }
\mathbf{q}_f=(\mathbf{u}_f,p_f)=(u_f,v_f,p_f),\ \mathbf{d}_f=(dx_f,dy_f),
\label{eq:state_fluid}
\end{equation}
and the structural-interface state as
\begin{equation}
S^{t}=\bigl(\mathbf{u}_s,\mathbf{d}_s\bigr)^{t},
\quad \text{where }
\mathbf{u}_s=(u_s,v_s),\ \mathbf{d}_s=(dx_s,dy_s).
\label{eq:state_struct}
\end{equation}

To couple the fluid and structure, a staggered (Z-type) partitioned procedure is employed, in which the structural interface state is advanced ahead of the fluid by one time step (Algorithm~\ref{alg:computation-process}). Specifically, the fluid prediction starts from state $F^{t}$, while the structural prediction starts from $S^{t+1}$. This design allows the fluid surrogate at time $t$ to operate on the updated structural interface geometry at $t+1$, which is essential in an ALE formulation since the deforming fluid mesh must be updated before advancing the flow solution.

Within each time step, the fluid mesh is first updated using the interfacial displacement from the structure at $t+1$ as:
\begin{equation}
\mathbf{d}_f^{t+1}=\mathcal{M}\!\left(\mathbf{d}_s^{t+1}\right),
\quad \text{with } 
\mathbf{d}_f^{t+1}\big|_{\Gamma_\text{beam}}=\mathbf{d}_s^{t+1}.
\label{eq:mesh_update}
\end{equation}
Here, $\mathcal{M}(\cdot)$ denotes mesh regeneration using Gmsh. To enforce the displacement continuity on the interface (Eq.~\eqref{eq:extension_bc}), the interface is discretized using the structural boundary nodes. These nodes are directly prescribed as the boundary nodes of the fluid mesh in Gmsh. As a result, the fluid and structural discretizations coincide at the interface, and thus, no interpolation between fluid and structural boundary nodes is required.

Given the updated mesh, the trained fluid surrogate model, denoted by $\mathcal{N}_f$, predicts the fluid state at $t+1$ from the current state as
\begin{equation}
\hat{\mathbf{q}}_f^{t+1}
=
\mathcal{N}_f\!\left(\mathbf{q}_f^{t},\,\mathbf{d}_f^{t},\,\mathbf{d}_f^{t+1},\,\mathbf{c}\right),
\label{eq:fluid_forward}
\end{equation}
where $\hat{\mathbf{q}}_f^{t+1}=(\hat{\mathbf{u}}_f,\hat{p}_f)^{t+1}$. The hat $(\hat{\cdot})$ denotes a predicted quantity, and $\mathbf{c}$ is a condition vector (illustrated in Figure~\ref{fig:vit-gno}) that encodes scenario information. Together with the mesh-update step in Eq.~\eqref{eq:mesh_update}, Eq.~\eqref{eq:fluid_forward} yields the predicted fluid state at the next time step $\hat{F}^{t+1}=(\hat{\mathbf{q}}_f,\mathbf{d}_f)^{t+1}$. A key challenge in learned partitioned coupling is the kinematic mismatch induced by staggered updates: the unconstrained fluid-side prediction may violate the interfacial kinematic compatibility in Eq.~\eqref{eq:kinematic_IC}. To enforce this constraint, we introduce an ALE-consistent interface velocity correction by imposing the structural interfacial velocity on the fluid interface,
\begin{equation}
\hat{\mathbf{u}}_f^{t+1}\big|_{\Gamma_\text{beam}} = \mathbf{u}_s^{t+1},
\label{eq:bc_correct}
\end{equation}
which enforces as a pointwise assignment on $\Gamma_\text{beam}$ (without additional interpolation). This treatment mitigates slip-like interfacial inconsistencies that would otherwise bias the predicted pressure and shear, degrade load transfer, and amplify errors over long rollouts.

Next, the flow-induced pressure on the wetted surface of the beam is then transferred to the structure model via
\begin{equation}
p_s^{t+1} = \mathcal\!\hat{p}_f^{t+1}\big|_{\Gamma_\text{beam}},
\label{eq:pressure_transfer}
\end{equation}
and the structural surrogate $\mathcal{N}_s$ updates the interfacial state,
\begin{equation}
\hat S^{t+2} = \mathcal{N}_s\!\left(\mathbf{u}_s^{t+1},\mathbf{d}_s^{t+1},\,p_s^{t+1},\,\mathbf{c}\right).
\label{eq:struct_forward}
\end{equation}

This completes one time-step advance of the coupled system, i.e., given $(F^{t},S^{t+1})$, the procedure predicts the next fluid state $\hat{F}^{t+1}$, incorporating the boundary-corrected velocity, and the structural state $\hat S^{t+2}$. Subsequent time steps are obtained by recursively repeating this procedure.

\begin{algorithm}[H]
  \caption{ALE-consistent GNO-Transformer framework for predicting FSI}
  \label{alg:computation-process}
  \small
  \begin{algorithmic}[1]
    \STATE \textbf{Initialization:} Provide initial fields and total rollout steps; the structure leads the fluid by one step.
    \STATE \hspace{1.2em} $F^{t}=(\mathbf{q}_f,\mathbf{d}_f)^{t}$ with $\mathbf{q}_f=(u_f,v_f,p_f)$, $\mathbf{d}_f=(dx_f,dy_f)$,
    \STATE \hspace{1.2em} $S^{t+1}=(\mathbf{u}_s,\mathbf{d}_s)^{t+1}$ with $\mathbf{u}_s=(u_s,v_s)$, $\mathbf{d}_s=(dx_s,dy_s)$,
    \STATE \hspace{1.2em} total rollout steps $T$.
    
    \FOR{$t = 0,1,\ldots,T-1$}
        \STATE \textbf{Mesh update:} Update the fluid mesh displacement using the structural boundary displacement.
        \STATE \hspace{1.2em} $\mathbf{d}_f^{t+1}\xleftarrow{\ \mathcal{M}\ }\mathbf{d}_s^{t+1} \qquad (\text{e.g., Gmsh})$
        
        \STATE \textbf{Fluid prediction:}
        \STATE \hspace{1.2em}
        \begin{tabular*}{\linewidth}{@{}l@{\extracolsep{\fill}}r@{}}
        $\hat{\mathbf{q}}_f^{t+1}
        \xleftarrow{\ \mathcal{N}_f\ }
        \bigl(\mathbf{q}_f^{t},\mathbf{d}_f^{t},\mathbf{d}_f^{t+1},\mathbf{c}\bigr)$
        & $\bm{\mathrm{loss}_f^{\,t+1}}$
        \end{tabular*}
        
        \STATE \textbf{ALE-consistent boundary correction:} Correct the fluid-side interface velocity by the structural interface velocity.
        \STATE \hspace{1.2em} $\tilde{\mathbf{u}}_f^{t+1}\big|_{\Gamma}\xleftarrow{}\mathbf{u}_s^{t+1},\qquad
        \tilde{\mathbf{q}}_f^{t+1}\leftarrow(\tilde{\mathbf{u}}_f,\hat{p}_f)^{t+1}$
        
        \STATE \textbf{Pressure transfer to structure:}
        \STATE \hspace{1.2em} $p_s^{t+1}\xleftarrow{ }\hat{p}_f^{t+1}\big|_{\Gamma_\text{beam}}$
        
        \STATE \textbf{Structure prediction:}
        \STATE \hspace{1.2em}
        \begin{tabular*}{\linewidth}{@{}l@{\extracolsep{\fill}}r@{}}
        $S^{t+2}
        \xleftarrow{\ \mathcal{N}_s\ }
        \bigl(S^{t+1},p_s^{t+1},\mathbf{c}\bigr)$
        & $\bm{\mathrm{loss}_s^{\,t+2}}$
        \end{tabular*}
        
        \STATE \textbf{State update:} $\mathbf{q}_f^{t+1}\leftarrow\tilde{\mathbf{q}}_f^{t+1}$,\quad $S^{t+2}\leftarrow(\mathbf{u}_s,\mathbf{d}_s)^{t+2}$.
    \ENDFOR
  \end{algorithmic}
\end{algorithm}

\subsection{Neural surrogates for fluid and structure}
\label{subsec:surrogates}

The fluid states are defined on deforming unstructured meshes and are therefore not directly compatible with image-based architectures such as CNNs or Vision Transformer (ViT)~\cite{dosovitskiy2020image}. To leverage Transformer-based spatiotemporal modeling while retaining the native unstructured discretization, we adopt a hybrid GNO-ViT-GNO pipeline (Figure~\ref{fig:vit-gno}, lower panel). In the coupling workflow described in Section~\ref{subsec:framework}, the fluid-predict operator $\mathcal{N}_f$ is implemented by this GNO-ViT-GNO surrogate.

The overall mapping from the current unstructured state to the next-step flow variables can be written as
\begin{equation}
\hat{\mathbf{q}}_f^{t+1}
=
\mathcal{P}_{\psi}\!\Bigl(
\mathcal{T}_{\omega}\!\bigl(\mathcal{L}_{\phi}(\mathbf{q}_f^{t},\mathbf{d}_f^{t},\mathbf{d}_f^{t+1}),\,\mathbf{c}\bigr)
\Bigr),
\label{eq:gno_vit_gno}
\end{equation}
where $\mathcal{L}_{\phi}$ is a  GNO that lifts unstructured node fields to a structured-grid tensor, $\mathcal{T}_{\omega}$ is a ViT module that performs global feature extraction and temporal advancement on patch tokens, and $\mathcal{P}_{\psi}$ is a second GNO that projects the structured prediction back to the native unstructured mesh.

Specifically, the lifting operators act on a mesh-induced graph
$G^t=(V^t,E^t)$, where $V^t=\{1,\dots,N_f\}$ indexes the $N_f$ nodes of the deforming fluid mesh at time $t$ with coordinates
$\mathbf{x}_i^t\in\mathbb{R}^2$, and $E^t$ contains undirected edges connecting mesh-neighboring nodes. Each node carries the feature vector
$\mathbf{f}_i^t=[\mathbf{q}_{f,i}^t,\mathbf{d}_{f,i}^t,\mathbf{d}_{f,i}^{t+1}]$:
\begin{equation}
\mathbf{X}^t=\mathcal{L}_{\phi}\bigl(\{\mathbf{x}_i^t,\mathbf{f}_i^t\}_{i\in V}\bigr),
\qquad
\hat{\mathbf{q}}_{f,i}^{t+1}=\mathcal{P}_{\psi}\bigl(\mathbf{Y}^{t+1},\mathbf{x}_i^{t+1}\bigr),\ \ i\in V,
\label{eq:lifting_projection}
\end{equation}
where $\mathbf{X}^t$ is the structured representation provided to the ViT and $\mathbf{Y}^{t+1}$ denotes the ViT-predicted structured output. Although the arguments of $\mathcal{N}_f$ in Eq.~\eqref{eq:fluid_forward} are written in terms of $(\mathbf{q}_f,\mathbf{d}_f)$, geometric information enters through the mesh graph $G=(V,E)$ and node coordinates $\mathbf{x}_i$ used by $\mathcal{L}_{\phi}$ and $\mathcal{P}_{\psi}$ in Eq.~\eqref{eq:lifting_projection}; hence the coordinate dependence of $\mathcal{N}_f$ is implicit.

The ViT stage operates on the structured tensor via patch tokenization. Let $\mathbf{X}^t$ be partitioned into $N$ patches $\{\mathbf{x}_k\}_{k=1}^{N}$ and embedded as
\begin{equation}
\mathbf{z}_k^0=\mathbf{W}_e\,\mathrm{vec}(\mathbf{x}_k)+\mathbf{e}_k,\qquad k=1,\ldots,N,
\label{eq:vit_tokens}
\end{equation}
where $\mathrm{vec}(\cdot)$ flattens each patch into a vector, $\mathbf{W}_e$ is a learnable linear embedding, and $\mathbf{e}_k$ is a positional embedding. Each Transformer block applies multi-head self-attention, providing a global receptive field that is well suited for capturing long-term interactions and maintaining temporal coherence in autoregressive rollouts. 
Let $\mathbf{Z}^0=[\mathbf{z}_1^0,\ldots,\mathbf{z}_N^0]\in\mathbb{R}^{N\times D_{emb}}$ denote the stacked patch tokens, where $D_{\mathrm{emb}}$ is the token embedding dimension. The ViT updates the tokens as
\begin{equation}
\mathbf{Z}^L=\mathcal{T}_{\omega}(\mathbf{Z}^0,\mathbf{c}),
\label{eq:vit_forward}
\end{equation}
and produces the structured prediction through a linear head followed by unpatchifying:
\begin{equation}
\mathbf{Y}^{t+1}=\mathrm{Unpatch}\!\left(\mathbf{W}_o\,\mathbf{Z}^L+\mathbf{b}_o\right),
\label{eq:vit_outputY}
\end{equation}
where $\mathbf{Y}^{t+1}$ has the same structured-grid format as $\mathbf{X}^t$ and is subsequently fed into the projection GNO $\mathcal{P}_{\psi}$ in Eq.~\eqref{eq:lifting_projection}. This design avoids mesh unfolding and preserves outputs on the evolving unstructured mesh required for near-wall fidelity.

\begin{figure}
    \centering
    \includegraphics[width=0.5\linewidth]{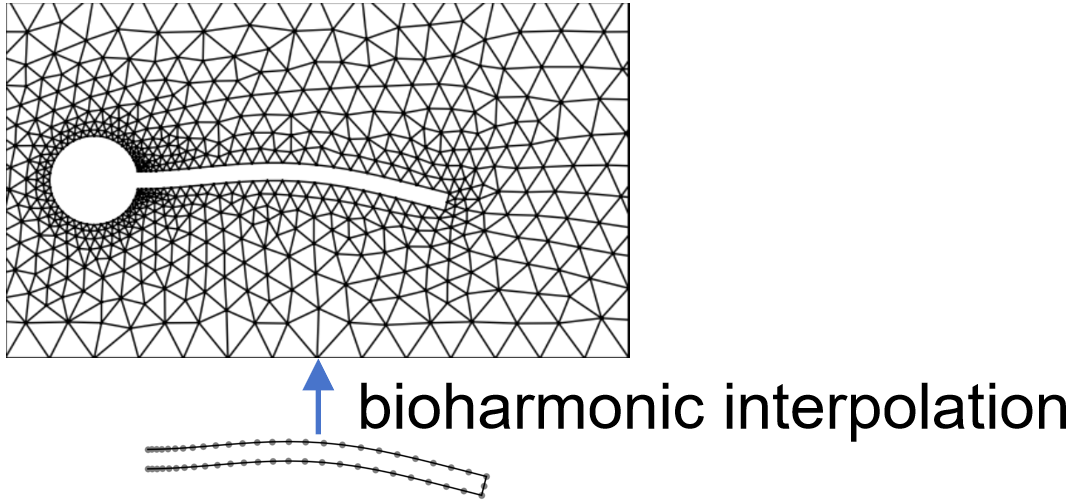}
    \caption{{Gmsh} generate fluid mesh based on the structure boundary.}
    \label{fig:Gmsh}
\end{figure}

For the structure, the surrogate predicts only the interfacial kinematics on $\Gamma_{\text{beam}}$. The structural update operator $\mathcal{N}_s$ in Eq.~\eqref{eq:struct_forward} is implemented by a lightweight LSTM~\cite{hochreiter1997long} applied to the ordered boundary-point sequence. We flatten the boundary-point kinematics into a 1D sequence using a vectorization operator $\mathcal{B}(\cdot)$ and predict the interface state as
\begin{equation}
\mathbf{s}^{t+2}
=
\mathrm{LSTM}\!\left(\mathbf{s}^{t+1},\,p_s^{t+1},\,\mathbf{c}\right),
\qquad
\mathbf{s}^{t}=\mathcal{B}\!\left(S^{t}\right),
\label{eq:lstm_update}
\end{equation}
which is reshaped back to obtain $S^{t+2}$. The predicted boundary serves three roles per time step: (i) it is used as the next interfacial state for autoregressive rollout, (ii) it determines the geometry for mesh update in Eq.~\eqref{eq:mesh_update}: regenerating the fluid mesh based on structure boundary (Figure~\ref{fig:Gmsh}), and (iii) it provides the interface velocity required by the boundary correction in Eqs.~\eqref{eq:bc_correct}.

\subsection{Two-stage training for stable long-term rollouts}
\label{subsec:training}

Long-term FSI prediction is performed autoregressively, which makes stable rollout sensitive to error accumulation. To address this, we propose a two-stage training protocol. The loss of different time step $t$ in Figure~\ref{fig:vit-gno} is the core of this method.
We first establish short-term prediction accuracy through single-step pretraining, and then fine-tune the coupled surrogate under an autoregressive rollout objective to explicitly suppress error accumulation under closed-loop feedback.

\paragraph{Stage I: Single-step pretraining (short-term)}
In the first stage, the fluid and structural models are trained for single-step prediction under teacher forcing. We use mean-squared errors evaluated on the fluid nodes and the structural boundary points, respectively:
\begin{equation}
\mathrm{loss}_f^{t+1}
=
\frac{1}{N_f}\left\|
\hat{\mathbf{q}}_f^{t+1}-\mathbf{q}_{f}^{t+1}
\right\|_2^2,
\qquad
\mathrm{loss}_s^{t+2}
=
\frac{1}{N_s}\left\|
\hat{\mathbf{s}}^{t+2}-\mathbf{s}_{}^{t+2}
\right\|_2^2,
\label{eq:one_step_losses}
\end{equation}
where $N_f$ denotes the number of fluid degrees of freedom used in the loss and $N_s$ denotes the number of boundary points (after vectorization). In this stage, GNO-ViT and LSTM are optimized separately using $\mathrm{loss}_f^{t+1}$ and $\mathrm{loss}_s^{t+2}$, respectively, yielding stable local predictors that provide a good initialization for long-term training.

\paragraph{Stage II: Long-term autoregressive fine-tuning}
In the second stage, we jointly fine-tune the fluid and structure models under an autoregressive rollout objective so that the coupled surrogate is explicitly trained on its own prediction feedback. Over a rollout window of length $n$, the rollout follows the same ALE-consistent coupling procedure as inference (Algorithm~\ref{alg:computation-process}), i.e., mesh update $\rightarrow$ fluid prediction $\rightarrow$ boundary correction $\rightarrow$ pressure transfer $\rightarrow$ structure prediction. The total loss is accumulated over the rollout window as
\begin{equation}
\mathrm{loss}_{\mathrm{total}} = \sum_{i=0}^{n} \left( \mathrm{loss}_{f}^{t+i} + \alpha \cdot \mathrm{loss}_{s}^{t+i+1} \right),
\label{eq:total_loss}
\end{equation}
where $\alpha=10$ balances the fluid and structure contributions. Due to GPU memory constraints, we set $n=5$ and apply truncated backpropagation through time over the rollout window. Unlike $Stage~I$, Eq.~\eqref{eq:total_loss} updates GNO-ViT and LSTM simultaneously, enabling the model to learn coupled long-term dependencies and to reduce cumulative error. Importantly, the ALE-consistent boundary correction is applied throughout the rollouts in this stage, aligning the training dynamics with inference-time coupling and preventing interface-induced kinematic inconsistencies from dominating the long-term loss.

\section{Results and Discussion}
\label{sec:Results}

% 在本文中，主要是对流固耦合预测过程中的流场预测进行了优化，固体只对边界点位移进行预测，维度较低，预测较为简单，限于篇幅，结果主要讨论都是对更复杂的流场预测的结果。特别地，将所提出的框架后续简写为GNO-ViT。
%在本部分中，第一节利用所提出的ViT-GNO框架预测流固耦合的有效性和准确性，分别对比了单步预测能力以及多步预测能力，以及其对边界信息的预测能力；其余三节为消融研究，第二节消融ViT，即GNO-ViT与纯GNO框架对比，第三节消融边界修正，即GNO-ViT与GNO-ViT-noBM（GNO-ViT-no boundary correction）分别对比了有无边界修正的边界预测能力以及长时预测能力；第四节则是消融长序列预测，对比基于长短序列训练(long-term)的模型的长时预测能力，并对比其累积误差。通过一系列消融与对比，说明提出的框架以及训练方法的高效性和物理一致性，以及各个模块的必要性及其作用，为后续工作提供指导。

This section presents a comprehensive evaluation of the proposed framework, including predictions in both periodic and non-periodic regimes of the benchmark FSI problem, as well as ablation studies.
%Because we target the high-dimensional problem of flow-field prediction within FSI. The analysis and discussion focus on the flow field, with structural results reported briefly for context. 

\subsection{Case settings}
\label{subsec:case-settings}
The benchmark FSI problem (Figure~\ref{fig:Fig-case}) involves the flexible beam vibration in the wake of a cylinder, and has been widely used for validation of FSI solvers~\cite{gao2024predicting, lee2024data, rahman2024pretraining}.
We follow the geometric configuration and material parameters reported in ~\cite{rahman2024pretraining}.
The computational domain is a channel of length $L=2.5~\mathrm{m}$ and height $H=0.41~\mathrm{m}$.
A rigid cylinder of diameter $d=0.1~\mathrm{m}$ is centered at $(0.2~\mathrm{m},\,0.2~\mathrm{m})$.
A flexible beam of height $h=0.02~\mathrm{m}$ and length $l=0.35~\mathrm{m}$ is attached to the rear side of the cylinder.
The rightmost beam tip (point A) located at $(0.6~\mathrm{m},\,0.2~\mathrm{m})$ is used as the monitoring point for structural motion.

Boundary conditions are prescribed as follows. At the inlet $\Gamma_{\mathrm{in}}$, a fourth-order velocity profile parameterized by $(c_1,c_2)$ is imposed.
Let $\bar{y}=y/H\in[0,1]$. The inlet condition is written as
\begin{equation}
u_{\mathrm{in}}(y)=U_{\infty}\,
\frac{11.7\bar{y}(1-\bar{y})(\bar{y}-c_1)(\bar{y}-c_2)}{(1-c_1)(1-c_2)},
\qquad \mathbf{u}_f=(u_{\mathrm{in}},0)\ \text{on } \Gamma_{\mathrm{in}},
\label{eq:inlet_bc}
\end{equation}
where $U_{\infty}=4~\mathrm{m/s}$. No-slip is applied on the top and bottom channel walls and on the cylinder boundaries
\begin{equation}
\mathbf{u}_f=\mathbf{0}, \quad \text{on } \Gamma_{\text{top}} \cup \Gamma_{\text{bottom}} \cup \Gamma_{\text{cwall}},
\label{eq:wall_bc}
\end{equation}
and a reference pressure is imposed at the outlet,
\begin{equation}
p_f=0, \quad \text{on } \Gamma_{\text{out}}.
\label{eq:outlet_bc}
\end{equation}
At the fluid-structure interface, the moving no-slip condition and traction equilibrium are enforced, consistent with Eqs.~\eqref{eq:kinematic_IC} and~\eqref{eq:dynamics_IC}.

For the material parameters, we adopt $\rho_f=1.0\times 10^{3}~\mathrm{kg/m^3}$, $\rho_s=1.0\times 10^{3}~\mathrm{kg/m^3}$,
$\lambda_s=4.0\times10^{6}~\mathrm{Pa}$, and $\mu_s=2.0\times10^{6}~\mathrm{Pa}$. 
With $U_{\infty}=4.0~\mathrm{m/s}$, $\mu_f=1.0~\mathrm{Pa\cdot s}$, all simulations are conducted at $Re=\frac{\rho_f U_\infty d}{\mu_f}=400$.

The dataset is constructed by varying the inlet profile in Eq.~\eqref{eq:inlet_bc}. With $c_1=-4.0$ fixed, we sample
$c_2\in\{-4.0,-2.0,0,2.0,4.0,6.0\}$. The subset $\{-4.0,0,2.0,4.0\}$ is used for training, while $\{-2.0,6.0\}$ is reserved for testing under interpolation and extrapolation settings, respectively. In addition, predictions are evaluated in both the
fully developed periodic regime and the underdeveloped transient (non-periodic) regime to assess long-term stability under
oscillatory and strongly transient FSI dynamics.

The training data are generated from high-fidelity FSI simulations, in which the fluid subproblem is solved using \texttt{OpenFOAM} and the structural subproblem using \texttt{solidii}, with the two solvers coupled via \texttt{preCICE}. To reduce training cost, the high-fidelity solution data are subsequently projected onto a substantially coarser mesh
to form the training dataset. Specifically, the original fields computed on a mesh with approximately $3\times10^{4}$ nodes are mapped to a reduced
representation with 1052 nodes. This projection significantly reduces GPU memory consumption during training while retaining the essential flow-structure
dynamics, thereby demonstrating the practical feasibility of the proposed framework. All training experiments are performed on an NVIDIA RTX A6000 GPU with 48 GB of memory.

\subsection{FSI prediction with GNO-ViT}
\label{subsec:gno-vit-results}
% 先写从发展段到周期段都可以实现比较准确的预测，相位对的可以，但是在发展区域的峰值有明显误差
% 云图说明预测的不错
% 具体的只对周期段预测，周期段非常准确，不管是x还是y方向，并且云图也更加准确，误差更小，从而证明周期段预测相对简单，再结合前面非周期段的结果进行分析，最终得出结论本文方法在保证周期段非常准的情况下对非周期段还达到了比较不错的准确度

The GNO-ViT framework is evaluated on the FSI benchmark problem across the transient non-periodic regime and the subsequent periodic regime. Here, the \emph{non-periodic regime} denotes the evolution from the initial stage through the transient growth stage until the beam settles into a stable periodic vibration (referred to as the \emph{periodic regime}). This configuration is phase-sensitive under autoregressive prediction, such that errors incurred during the transient stage can propagate into the periodic stage. Therefore, accurate prediction in this setting requires not only capturing the transient dynamics but also maintaining phase consistency across regimes, making it a challenging test for long-term FSI prediction.

\begin{figure}
    \centering
    \includegraphics[width=1\linewidth]{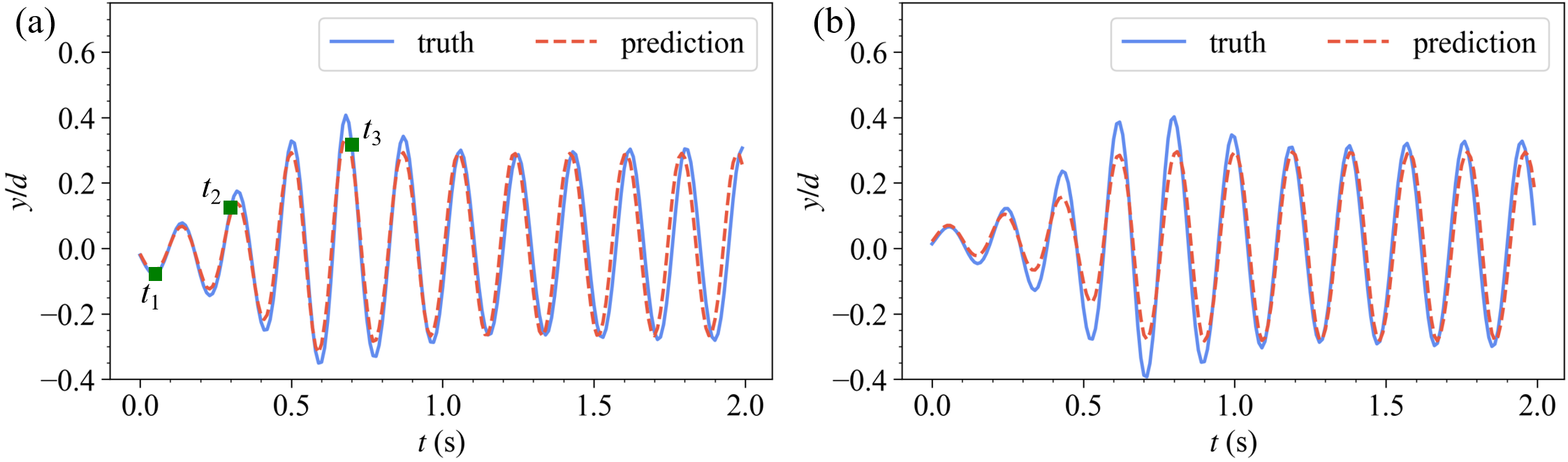}
    \caption{Prediction and truth for the time histories of normalized transverse displacement at monitoring point A in the non-periodic regime: (a) $y/d$ for $c_2=-2$; (b) $y/d$ for $c_2=6$.}
    \label{fig:non-per-line}
\end{figure}

Figure~\ref{fig:non-per-line} shows the normalized displacement histories of $y$-direction at probe A for $c_2=-2$ (interpolation) and $c_2=6$ (extrapolation). Across both cases, the proposed framework accurately captures the displacement evolution from the initial low-amplitude oscillations, through the transient non-periodic regime, and into the periodic regime. The phase of the predicted vibration matches well with the reference over the entire time, demonstrating stable rollouts over the entire evolution. Noticeable discrepancies are primarily confined to the transient growth stage, where coupled forcing varies rapidly and the energy transfer of wake structure is strongly unsteady. In this regime, small differences in vortex-shedding onset and wake adjustment can lead to deviations in instantaneous load. This results in local amplitude mismatch while preserving the overall phase trajectory.

\begin{figure}
    \centering
    \includegraphics[width=1\linewidth]{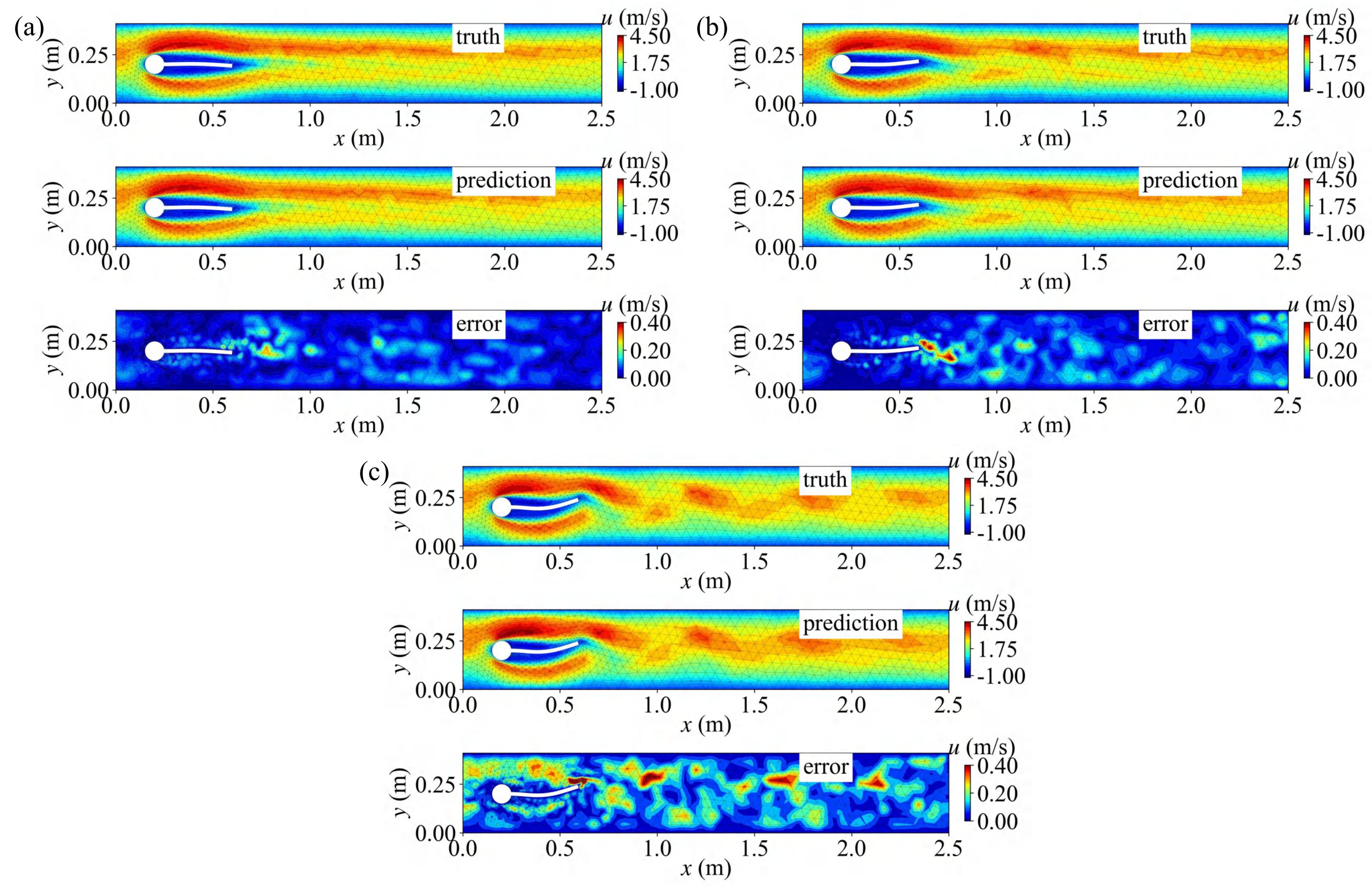}
    \caption{Comparison of prediction and truth of flow field $u_f$ on non-periodic regime for $c_2$=-2 at time (a) $t_1$, (b) $t_2$, (c) $t_3$.}
    \label{fig:non-per-con}
\end{figure}

Flow-field snapshots further support the above observations. Figure~\ref{fig:non-per-con} compares the predicted and reference $u_f$ at representative instants ($t_1,t_2,t_3$ shown in Figure~\ref{fig:non-per-line}(a)) for the non-periodic case $c_2=-2$. The framework accurately predicts the evolution of key flow features, including shear-layer morphology and vortex spacing, throughout the transient process. As the flow becomes strongly unsteady, discrepancies become more visible but remain largely localized near the trailing edge and in the near-wake region, where strong shear and nonlinear interactions dominate.

\begin{figure}
    \centering
    \includegraphics[width=1\linewidth]{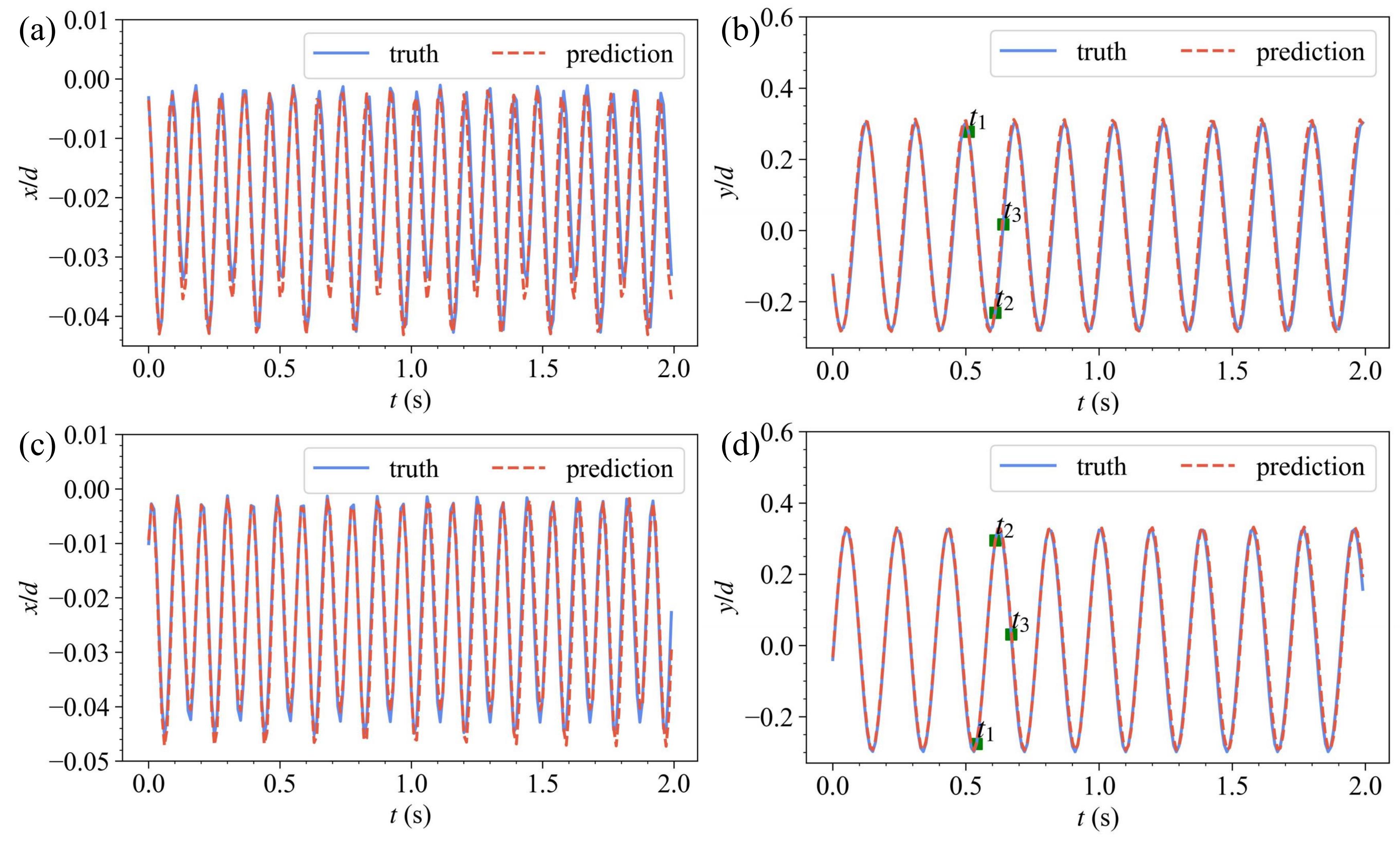}
    \caption{Prediction and truth for the time histories of normalized structural displacements at monitoring point A in the periodic regime: (a) $x/d$ for $c_2=-2$; (b) $y/d$ for $c_2=-2$; (c) $x/d$ for $c_2=6$; (d) $y/d$ for $c_2=6$.}
    \label{fig:disp-xy-2-6}
\end{figure}

Figure~\ref{fig:disp-xy-2-6} presents the normalized displacement histories in the periodic regime for both the $x$ and $y$-direction for $c_2=-2$ and $c_2=6$. In this regime, the predicted displacement show close agreement with the reference solution in both phase and amplitude, with smaller deviations than those observed during transient growth in Figure~\ref{fig:non-per-line}. 

\begin{figure}
    \centering
    \includegraphics[width=1\linewidth]{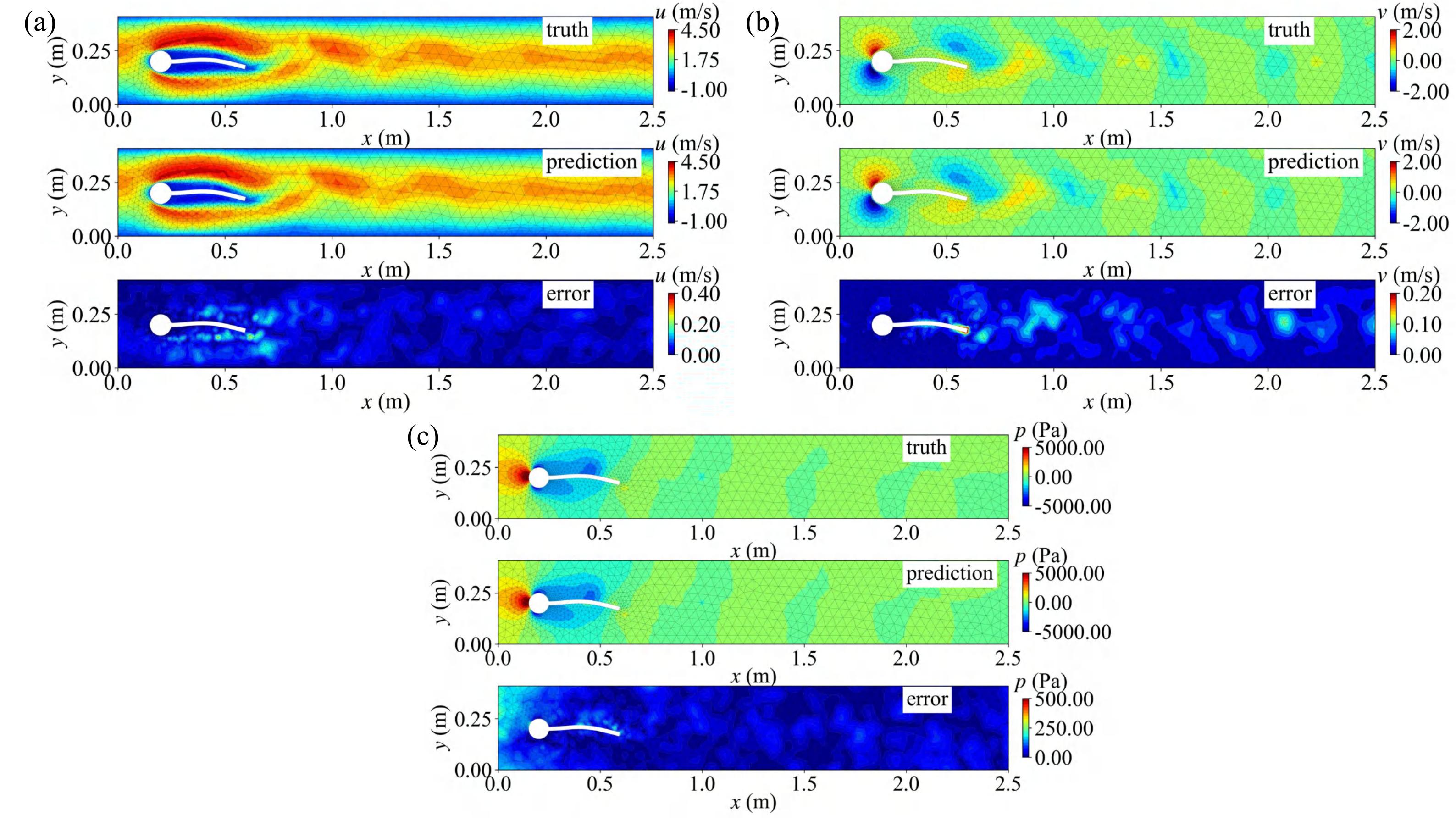}
    \caption{Comparison of prediction and truth of flow field at time $t_2$ for $c_2$=-2: (a) $u_f$, (b) $v_f$, (c) $p_f$.}
    \label{fig:contour-t2-x2}
\end{figure}

To further examine the flow-field prediction in the periodic regime, three representative instants ($t_1$, $t_2$, and $t_3$) are selected along the $y/d$ trajectories in Figure~\ref{fig:disp-xy-2-6} (b) and (d), corresponding to different phases of the beam oscillation. Figure~\ref{fig:contour-t2-x2} shows the predicted flow fields $(u_f,v_f,p_f)$ at time $t_2$ for the case $c_2=-2$.
Consistent with the displacement trend, the periodic flow-field comparison at time $t_2$ (Fig.~\ref{fig:contour-t2-x2}) shows higher fidelity and lower-magnitude, more localized errors for $(u_f,v_f,p_f)$, indicating that the stabilized periodic dynamics are comparatively easier to predict than the transient development stage. Additional results for the predicted flow fields at $t_1$ and $t_3$ for $c_2=-2$, as well as the corresponding comparisons for the extrapolative case $c_2=6$, are provided in ~\ref{sec:appendix} to further demonstrate generalization under both interpolation and extrapolation in the periodic regime. Supplementary Videos S1-S2 provide time-resolved animations comparing the predicted and reference velocity field $u_f$ together with the corresponding structural deformation for $c_2=-2$ and $c_2=6$ in periodic regime. The files are available via Zenodo~\cite{zhao2026_fsi_movies_zenodo}.

Overall, the proposed framework achieves highly accurate predictions in the periodic regime while retaining strong predictive capability during transient growth. Crucially, phase fidelity is maintained from the non-periodic development stage into the stable limit cycle, supporting reliable long-term autoregressive rollouts; the remaining transient errors predominantly manifest as localized peak-amplitude mismatch rather than global phase drift.

\subsection{Ablation studies}
\label{subsec:Ablation study}
This section presents ablation studies to quantify the contributions of key components, including the ViT module, the ALE-consistent boundary correction, and the long-term training strategy.

\subsubsection{Role of ViT module and ALE-consistent boundary correction}
\label{subsubsec:GNO-vit GNO}

To assess the respective roles of the ViT module and the ALE-consistent boundary correction, two ablated variants of the proposed framework are considered. First, to isolate the contribution of ViT module, the fluid prediction step in Algorithm~\ref{alg:computation-process} is modified by replacing the GNO-ViT block with a standalone GNO. Second, to evaluate the role of boundary correction, the ALE-consistent boundary correction step in Algorithm~\ref{alg:computation-process} is removed, such that the fluid velocity predicted by GNO-ViT is used directly without enforcing kinematic condition at the interface. This variant is referred to as GNO-ViT-noBM.

To provide a controlled evaluation setting, we consider a one-way configuration of the benchmark problem in which the structural motion is prescribed and drives the fluid prediction. Specifically, the structural displacement history on $\Gamma_{\text{beam}}$, together with its time derivative, is extracted from the high-fidelity FSI solution and imposed throughout the rollout. The fluid mesh and ALE mapping are updated accordingly at each time step. All ablation results are reported for an interpolative generalization case at $c_2=-2.0$, using the high-fidelity FSI solution as the truth. We report both instantaneous flow-field errors and the error growth during autoregressive rollouts over the entire evaluation window.

\begin{figure}[!t]
    \centering
    \includegraphics[width=1\linewidth]{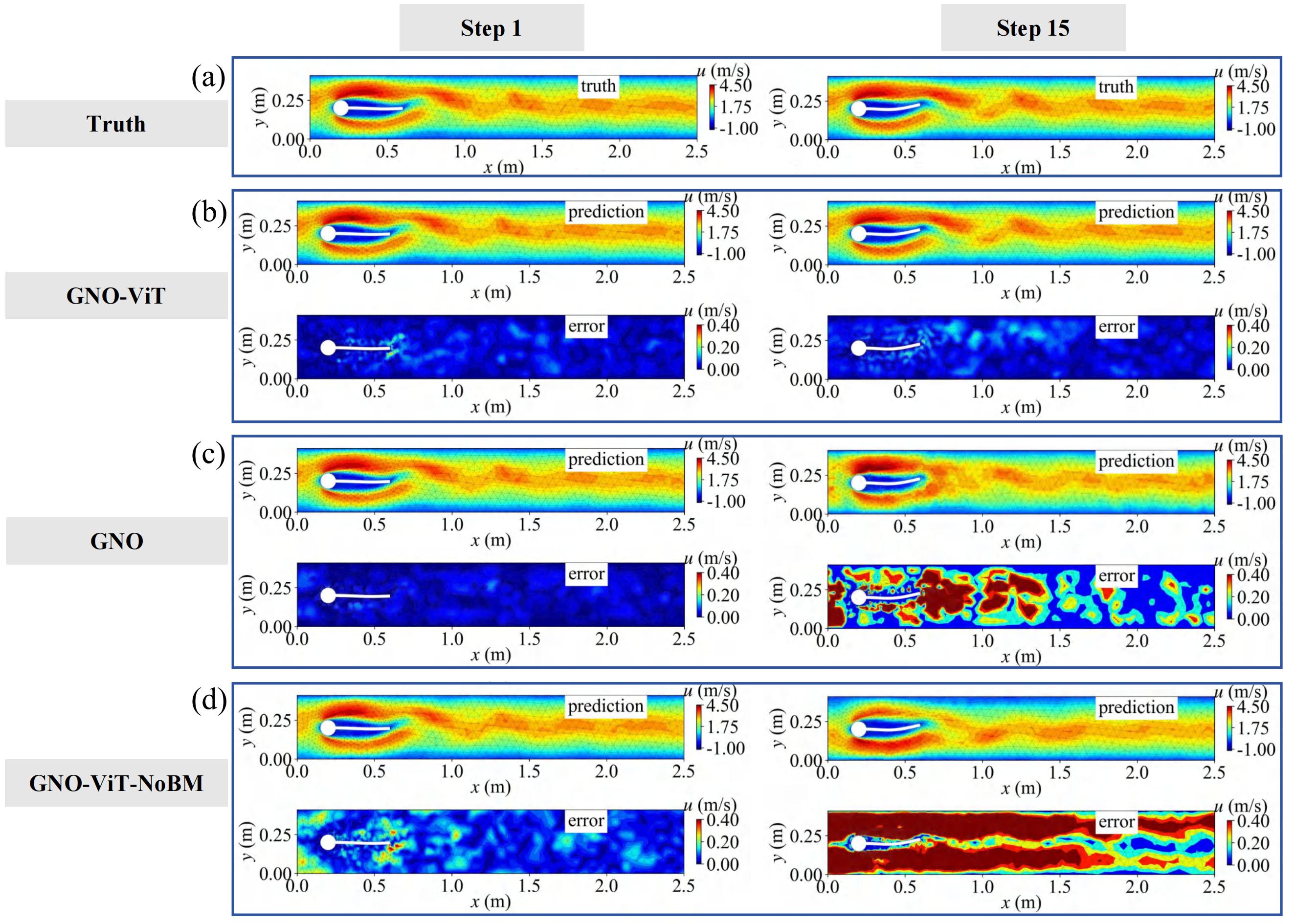}
    \caption{Truth and prediction results of flow field by different methods at step 1 and step 15 for $c_2$=-2: (a) truth, (b) GNO-ViT, (c) GNO, (d) GNO-ViT-noBM.}
    \label{fig:GNO-ViT vs GNO contour}
\end{figure}

Figure~\ref{fig:GNO-ViT vs GNO contour} compares the predicted streamwise velocity $u_f$ from GNO-ViT, pure GNO, and GNO-ViT-noBM at autoregressive steps 1 and 15. At step 1, pure GNO yields slightly smaller errors, indicating stronger single-step regression. However, its predictions rapidly lose coherence as the rollout proceeds, whereas GNO-ViT preserves the main flow structures and phase alignment at step 15. In contrast, GNO-ViT-noBM exhibits noticeable near-boundary discrepancies already at step 1 (especially around the beam and inflow), which amplify dramatically by step 15, consistent with the accumulation of interface-induced slip errors when kinematic enforcement is removed. Overall, the qualitative comparisons suggest that ViT is critical for maintaining coherent long-term evolution, while boundary correction is essential for preventing near-wall bias from destabilizing the rollout.

\begin{figure}
    \centering
    \includegraphics[width=1.0\linewidth]{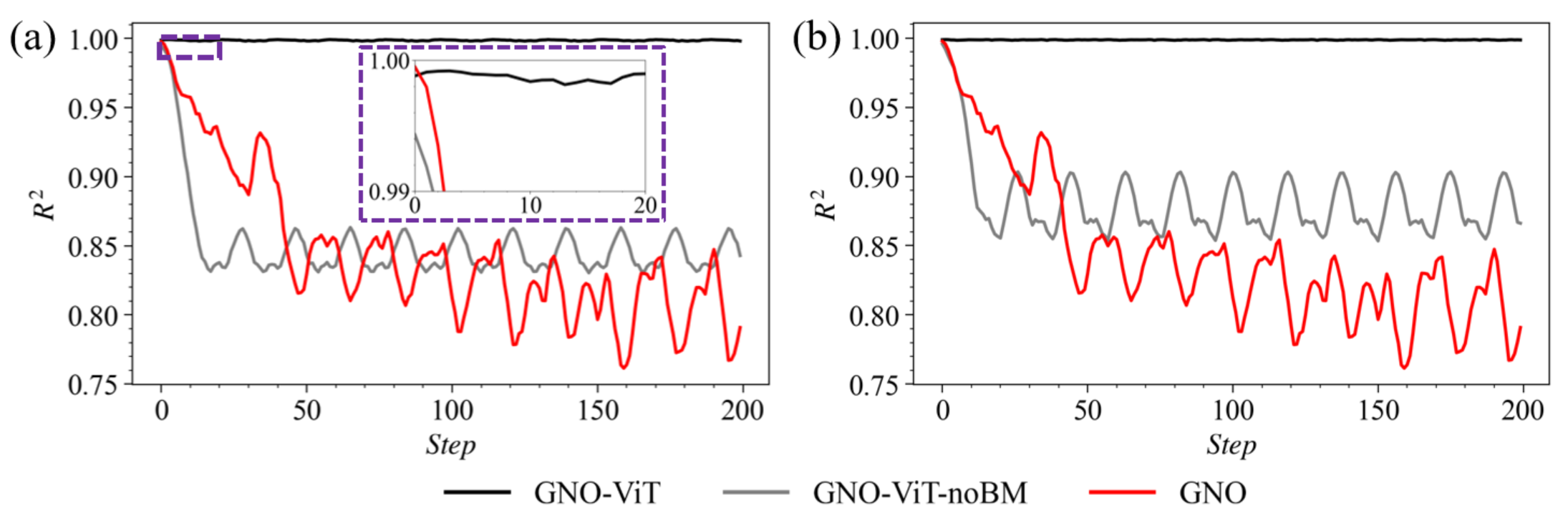}
    \caption{Time evolution of the correlation coefficient $R^2$ during rollouts for (a) velocity field $u_f$ and (b) the pressure field $p_f$, comparing GNO-ViT, GNO-ViT-noBM, and the GNO. The inset in (a) highlights the early-stage  of $u_f$.}

    \label{fig:GNO-ViT vs GNO R}
\end{figure}

Figure~\ref{fig:GNO-ViT vs GNO R} shows the temporal evolution of the flow-field correlation coefficient $R^2$ for the fluid velocity $u_f$ and $p_f$. For the Figure~\ref{fig:GNO-ViT vs GNO R}(a), the GNO exhibits a slightly higher $R^2$ initially but then decays rapidly, whereas GNO-ViT maintains an $R^2$ value close to unity throughout the rollout process. When the boundary correction is removed, $R^2$ drops rapidly at the beginning to approximately 0.85 and subsequently fluctuates around this level, indicating that temporal modeling alone cannot compensate for persistent near-interface inconsistencies. 

\begin{figure}
    \centering
    \includegraphics[width=1.0\linewidth]{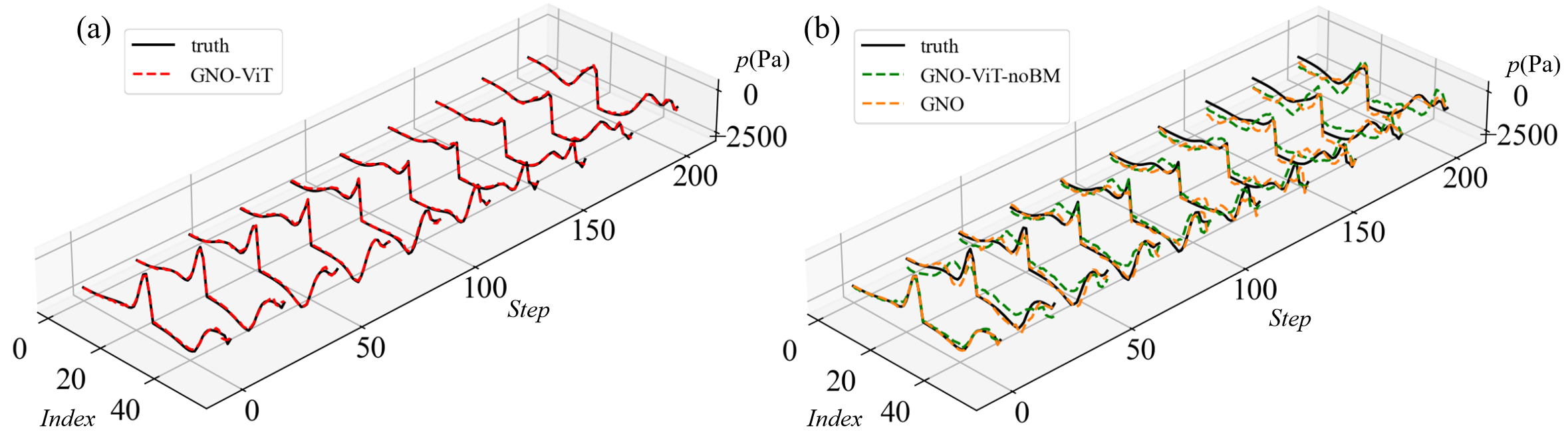}
    \caption{Comparison of boundary pressure predictions by GNO-ViT, GNO and GNO-ViT-NoBM for $c_2 = -2$: (a) GNO-ViT, (b) GNO and GNO-ViT-NoBM.}
    \label{fig:GNO-ViT GNO-ViT-NoBM bound p}
\end{figure}

\begin{figure}
    \centering
    \includegraphics[width=1\linewidth]{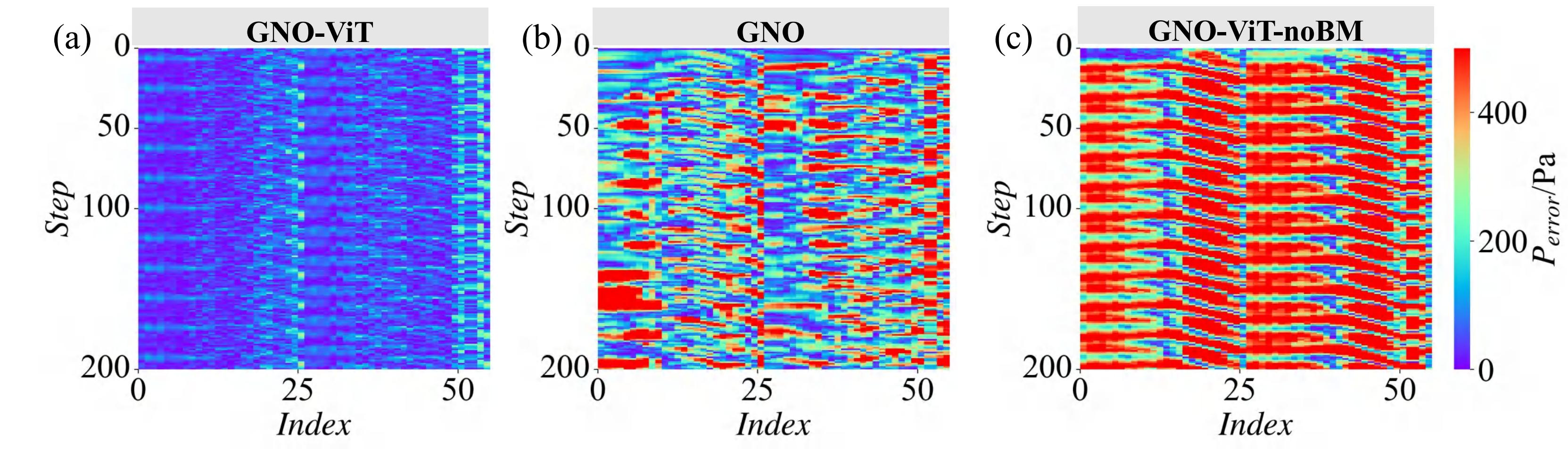}
    \caption{Time-evolution of predicted $p_f$ errors by different method for $c_2 = -2$: (a) GNO-ViT, (b) GNO, (c) GNO-ViT-noBM.}
    \label{fig:GNO-ViT vs GNO bound p}
\end{figure}

\begin{figure}
    \centering
    \includegraphics[width=0.6\linewidth]{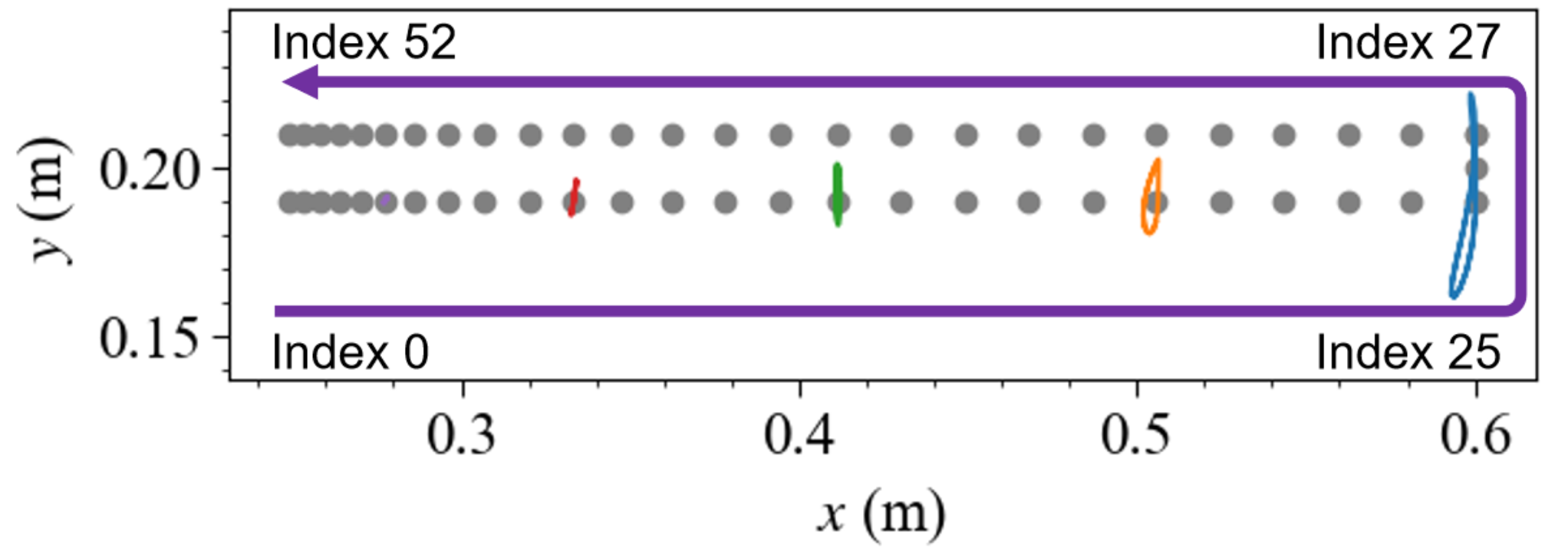}
    \caption{Structural boundary node indexing and representative displacements of the flexible beam in the $(x,y)$ plane. The boundary nodes are ordered from the beam root to the tip: indices 0-25 correspond to the lower edge and indices 27-52 correspond to the upper edge.}
    \label{fig:displacement beam nodes}
\end{figure}

Figures~\ref{fig:GNO-ViT GNO-ViT-NoBM bound p} and~\ref{fig:GNO-ViT vs GNO bound p} provide complementary comparisons of the predicted pressure $p_f$ along the flexible-beam boundary ($\Gamma_{\text{beam}}$ in Figure~\ref{fig:Fig-case}) and the corresponding error distributions. The $Index$ axis corresponds to the numbering of the sampled boundary points on the flexible beam shown in Figure~\ref{fig:displacement beam nodes}: the nodes are indexed counterclockwise along the beam boundary starting from the lower-left end and ending at the upper-left end, with $Index$ ranging from 0 to 53. The $Index$ used in the remainder of the paper has the same meaning as defined here.

Figure~\ref{fig:GNO-ViT GNO-ViT-NoBM bound p} compares the boundary-pressure histories predicted by GNO-ViT, GNO and GNO-ViT-NoBM for $c_2=-2$ about 200 steps. 
As shown in Figure~\ref{fig:GNO-ViT GNO-ViT-NoBM bound p}(a), the proposed GNO--ViT closely matches the ground truth across all sampled boundary locations, accurately predicting both the peak amplitudes and the timing of rapid pressure transients.
Figure~\ref{fig:GNO-ViT GNO-ViT-NoBM bound p}(b) reports the ablation results. Although GNO and GNO-ViT-NoBM remain close to the reference at early steps, both exhibit a progressive deterioration as the rollout proceeds, characterized by attenuated pressure peaks and accumulated phase drift.
Among them, GNO-ViT-NoBM shows the most pronounced deviation, indicating that removing the boundary-correction constraint leads to a significant loss of near-boundary fidelity and long-horizon stability.

Figure~\ref{fig:GNO-ViT vs GNO bound p} shows time-resolved error maps along the beam boundary. For GNO-ViT, the map remains largely blue with thin, periodic bands that follow the shedding frequency, indicating bounded, non-accumulating errors. For GNO, alternating red/blue bands appear and grow with time. This is indicative of phase drift during autoregressive prediction. In the absence of explicit temporal modeling, local graph convolutions fit each step well but fail to maintain the phase of coherent structures across steps, so the errors oscillate and amplify. For GNO-ViT-noBM, the map is dominated by red color with stripe-like patterns that persist over time. This reflects a systematic bias caused by violating the interface kinematics. The resulting slip at the wall perturbs the near-boundary shear which induces a sustained pressure offset rather than random noise, hence the saturated error levels. This observation is also confirmed by the time evolution of the correlation coefficient $R^2$ for the fluid pressure field $p_f$ (Figure~\ref{fig:GNO-ViT vs GNO R}(b)).

Besides, the difference between GNO and GNO-ViT-noBM in Figure~\ref{fig:GNO-ViT vs GNO bound p} and Figure~\ref{fig:GNO-ViT vs GNO R}(b) reveals an noteworthy contrast between global flow accuracy and boundary pressure accuracy. Toward the end of the rollout, the $R^2$ of the global fluid pressure field drops to roughly 0.75 for GNO and 0.85 for GNO-ViT-noBM, yet the structural boundary pressure error of GNO is still smaller than GNO-ViT-noBM. This outcome underscores the necessity and efficiency of the boundary correction method: even when the overall flow prediction deteriorates, enforcing the modified boundary velocities preserves higher fidelity in the boundary pressure prediction, confirming that accurate interface prediction depends more on enforcing kinematic fidelity than on global flow correlation.

\begin{figure}
    \centering
    \includegraphics[width=1\linewidth]{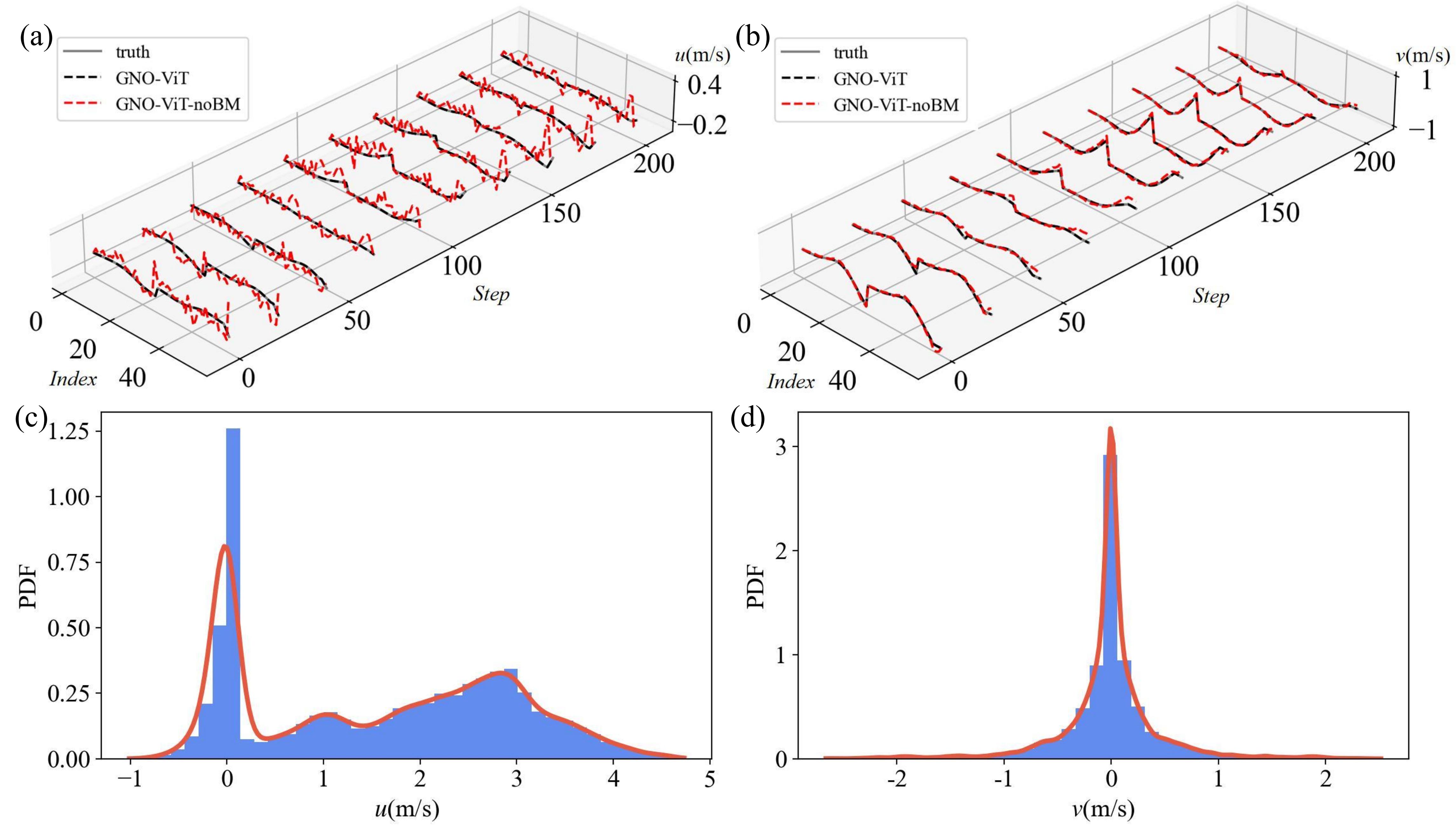}
    \caption{Prediction and distribution of structural boundary velocity for GNO-ViT and GNO-ViT-NoBM: predicted (a) $u_f$, (b) $v_f$, distribution of original (c) $u_f$, (d) $v_f$.}
    \label{fig:GNO-ViT GNO-ViT-NoBM bound uv}
\end{figure}

Figure~\ref{fig:GNO-ViT GNO-ViT-NoBM bound uv} compares the predicted interface velocities obtained with GNO-ViT and GNO-ViT-noBM. Panels (a) and (c) show the streamwise velocity $u_f$, while panels (b) and (d) show the transverse velocity $v_f$. Owing to the ALE-consistent boundary correction, the GNO-ViT prediction matches well with the reference. In contrast, GNO-ViT-noBM yields a noisy and biased prediction for $u_f$, whereas its prediction for $v_f$ remains relatively closer to the reference.
The different behaviors of $u_f$ and $v_f$ can be attributed to both spatial complexity and magnitude disparity. As illustrated by the distributions in Figures~\ref{fig:GNO-ViT GNO-ViT-NoBM bound uv}(c) and (d), $u_f$ exhibits stronger local gradients and more pronounced sign variations, making it more difficult to learn. Moveover, the transverse interface motion is substantially larger in magnitude than the streamwise motion along the beam, making $u_f$ more susceptible to noise and regression errors. Even after normalization, $u_f$ remains small relative to the main flow velocity scale, which further increases the difficulty of stable learning. Since boundary data directly mediate the FSI coupling, inaccurate interface velocities propagate to near-wall shear, pressure distribution, and load transfer. The boundary correction replaces the fluid-side boundary velocity with the physically consistent structural velocity. This constraint effectively suppresses interface drift during rollouts and guides the surrogate toward stable and physically admissible predictions.

Overall, these results demonstrate the role of temporal coherence (enabled by the ViT’s global self-attention) and spatial boundary consistency (enforced by ALE-consistent boundary correction) in the proposed framework. The former helps maintain phase-locked evolution of coherent vortical structures and suppresses error propagation during autoregressive rollouts, while the latter enforces interfacial kinematic compatibility and mitigates slip-induced near-wall pressure bias.

%%%%%%%%%%%%%%%%%%%%%%%%%%%%%%%%%%%%%%%%%%%%%%%%%%%%%%%%%%%%%
\subsubsection{Role of long-term training}
\label{subsubsec:seq2seq}
% 前述章节主要讨论了网络结构配置的影响；本节则聚焦训练策略，通过消融长时序训练机制来评估其作用。在该消融实验中，流体网络与结构网络仅采用单步监督分别训练，随后直接进行耦合用于 FSI 预测，而不在训练过程中累积多个未来时间步的滚动（rollout）损失。

% 图~\ref{fig:short-term FSI} 给出了在 (c_2=-2) 与 (c_2=6) 两种入口条件下，采用单步监督分别训练后直接耦合得到的模型，对图~\ref{fig:Fig-case} 中柔性梁测点 A 的 (x) 与 (y) 向位移预测结果及真实值对比。可以看到，该“短期训练”模型在预测初期能够较高精度复现参考信号的周期振荡特征。然而，随着时间推进，预测轨迹逐渐出现持续累积的相位滞后，这是典型的时间漂移误差累积现象。其根源在于单步监督范式仅最小化瞬时误差，并未显式考虑在长时序滚动推演中“预测作为下一步输入”所引发的误差传播机制。因此，早期微小的局部偏差会在自回归过程中不断放大，最终导致位移波形与真实值逐步失同步。

% 图~\ref{fig:long-short-term FSI R} 中相关系数 (R^2) 的时间演化进一步验证了上述退化趋势。对于 (c_2=-2) 与 (c_2=6) 两种情况，短期训练模型（红色）的 (R^2) 随时间显著下降，并伴随准周期振荡；相比之下，长时序训练模型（黑色）能够在整个推演过程中保持稳定且较高的 (R^2)，表明其时间鲁棒性更强。长时序训练通过引入自回归一致性约束，使模型在训练阶段显式经历自身预测反馈，从而能够“自校正”并内化跨时间依赖关系，进而形成对底层物理动力学更稳定的表征。

% 当前的长时序训练策略在概念上与“一次性多步预测”（one-shot multi-step predicting）相近，二者均旨在缓解未来多步误差累积。然而，在 one-shot 设置下，模型在一次更新中接收聚合的多步误差信号；而长时序训练则通过自回归滚动推演逐步获得这些误差。尽管两者在优化逻辑上具有相似性，但长时序训练在面向流动控制任务时具有关键优势：由于 FSI 场是逐步生成的，控制输入可以在任意时间步灵活施加；而 one-shot 生成缺乏这种逐步交互的能力，因此不适用于此类控制场景。

% 总体而言，结果强调：单步监督能够保证局部短期精度，而长时序训练则能够培养全局时间相干性与抗误差累积能力，使 FSI 动力学在更长预测时域内保持物理一致性，这对于高保真 FSI 仿真与闭环控制应用至关重要。

To examine the role of long-term training strategy, we conduct an ablation in which the fluid and structure models are trained separately with single-step (short-term) supervision and are then directly coupled for FSI prediction, without accumulating rollout losses over multiple future steps.

\begin{figure}
    \centering
    \includegraphics[width=1\linewidth]{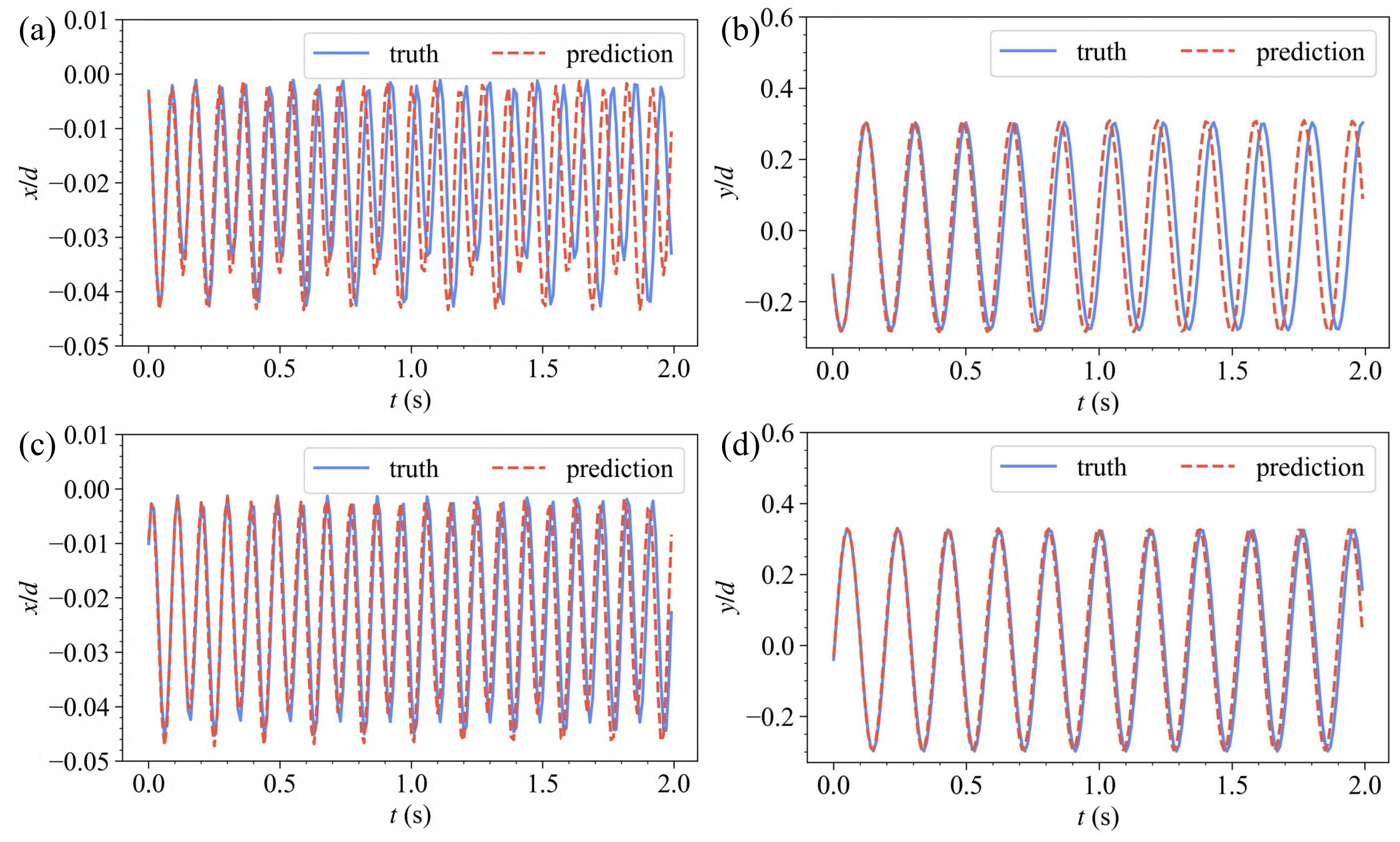}
        \caption{Prediction and truth for the time histories of normalized structural displacements at monitoring point A in the periodic regime using short-term training only (without long-term fine-tuning): (a) $x/d$ for $c_2=-2$; (b) $y/d$ for $c_2=-2$; (c) $x/d$ for $c_2=6$; (d) $y/d$ for $c_2=6$.}
    \label{fig:short-term FSI}
\end{figure}

Figure~\ref{fig:short-term FSI} compares the predicted and reference for $x$ and $y$ displacement history at point A for $c_2=-2$ and $c_2=6$ obtained without long-term training. The short-term trained model reproduces the periodic response with high accuracy at early time. However, as time advances, the predicted trajectory develops a progressive phase lag relative to the reference solution, indicating cumulative temporal drift. This behavior originates from the one-step supervision paradigm, in which training minimizes only instantaneous losses without accounting for error propagation that occur when predictions are recursively fed back as inputs during long rollouts. As a result, minor local discrepancies amplify over time, leading to progressive desynchronization of the displacement waveform.

\begin{figure}
    \centering
    \includegraphics[width=1\linewidth]{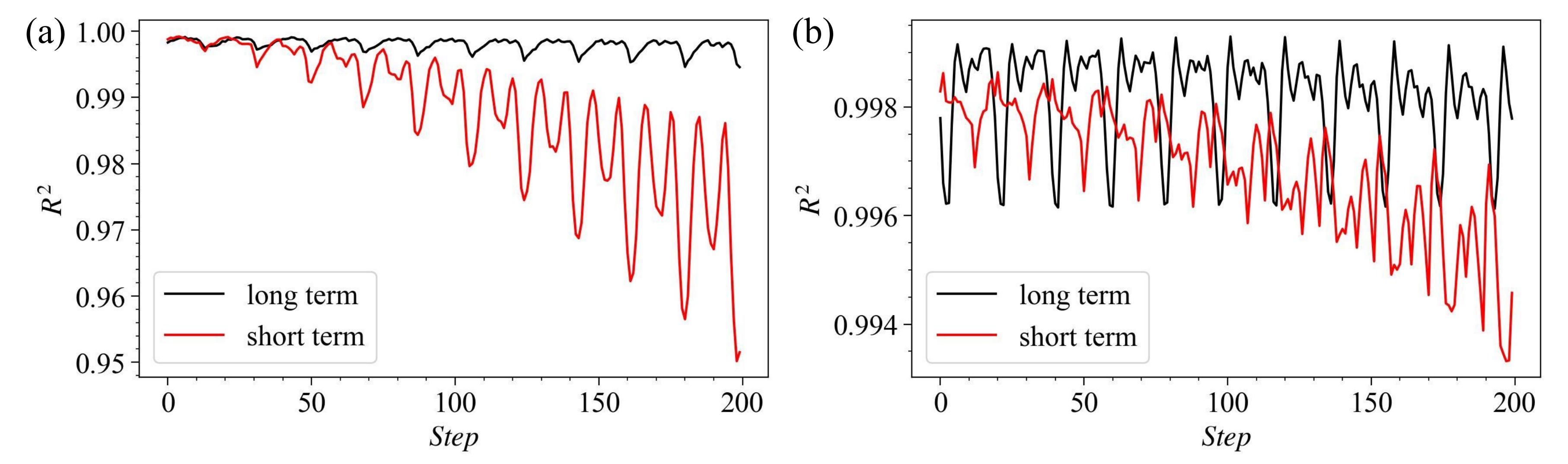}
    \caption{Time evolution of the flow-field correlation coefficient $R^2$ of $u_f$ between prediction and truth for short-term and long-term training: (a) $c_2=-2$; (b) $c_2=6$.}

    \label{fig:long-short-term FSI R}
\end{figure}

The degradation is further quantified by the temporal evolution of the flow-field correlation coefficient $R^2$, as shown in Figure~\ref{fig:long-short-term FSI R}. 
For both $c_2=-2$ and $c_2=6$, the short-term trained model shows a clear decrease in $R^2$ as the rollout progresses, accompanied by quasi-periodic flucturations. In contrast, the long-term trained model consistently maintains high $R^2$ values, indicating improved temporal robustness. This difference highlights the importance of long-term training, which introduces autoregressive consistency constraints by explicitly exposing the model to its own prediction feedback during training. This self-supervised correction enables it to internalize the temporal dependencies and develop a more stable representation of the underlying physical dynamics.

The proposed long-term training strategy is conceptually similar to one-shot multi-step predicting in that both aim to mitigate error accumulation across future steps. The key difference lies in how multi-step error is exposed. In the one-shot setting, the model receives the aggregate multi-step error in a single update, whereas long-term training obtains these errors progressively through autoregressive rollouts. Although the optimization logic is comparable, the above difference is crucial for the flow-control applications. Because the FSI field is generated step-by-step, control inputs can be deployed at any desired time step. This flexibility is absent in one-shot generation, making it unsuitable for such control tasks.

Overall, while short-term supervision yields accurate one-step predictions, long-term training promotes global temporal coherence and error resilience. The latter ensures that the FSI dynamics remain physically consistent over extended prediction terms, which is essential for high-fidelity FSI simulation and closed-loop control.

% 下图是长短时训练模型

%%%%%%%%%%%%%%%%%%%%%%%%%%%%%%%%%%%%%%%%%%%%%%%%%%%%%%%%%%%%%%%%%%%%%%%%%%%
\section{Conclusions}
\label{sec:Conclusions}
This study proposed an ALE-consistent GNO-Transformer framework for FSI prediction on deforming unstructured meshes. The framework couples (i) a GNO-ViT surrogate for predicting the fluid state on deforming meshes, (ii) a lightweight LSTM for predicting interfacial structural kinematics, and (iii) an ALE-consistent boundary correction strategy that explicitly enforces interfacial kinematic condition. To improve rollout robustness, we further adopt a two-stage training protocol consisting of single-step pretraining followed by long-term autoregressive fine-tuning, which exposes the model to its own feedback and mitigates cumulative error growth.

Comprehensive evaluations using an FSI benchmark demonstrate accurate and stable prediction in both periodic and non-periodic (transient) regimes, as well as strong interpolation/extrapolation generalization under inlet-profile variations. Quantitatively, the proposed GNO-ViT framework maintains a flow-field correlation coefficient $R^2$ close to unity throughout the rollout. Besides, via ablation studies, we find that removing the ALE-consistent boundary correction causes an early drop of $R^2\approx 0.85$ for both $u_f$ and $p_f$, indicating that temporal modeling alone is insufficient to offset persistent interface inconsistencies. Moreover, replacing GNO-ViT with a pure GNO further degrades $R^2$ to $0.75$ for both $u_f$ and $p_f$ at the end of the rollout, further highlighting the role of global temporal modeling and interfacial kinematic enforcement in stabilizing long-term predictions. In addition, long-term autoregressive fine-tuning improves rollout stability by increasing the end-of-term $R^2$ from $0.95$ (single-step only) to $0.99$, which is consistent with the observed suppression of phase drift under recursive prediction. Beyond the global metrics, the ALE-consistent boundary correction also improves near-boundary fidelity: the boundary pressure predictions from GNO-ViT closely match the reference throughout the rollout, whereas removing the boundary correction introduces systematic near-wall bias and amplifies errors.

These results clarify the complementary roles of the four key components in the proposed framework: the GNO provides geometry-aware spatial modeling on unstructured domains; the ViT promotes long-range temporal coherence and mitigates phase drift during autoregressive rollouts; the long-term autoregressive training strategy suppresses compounding error accumulation and stabilizes multi-step predictions; and the ALE-consistent boundary correction enforces kinematic fidelity at the moving interface, preventing slip-induced near-wall bias from destabilizing the coupled feedback loop. 

Limitations remain in that the present study focuses on a canonical 2D benchmark. Future work will extend the framework to 3D FSI configurations, incorporate additional physics constraints during training, and explore tighter integration with high-fidelity ALE solvers for hybrid data-physics coupling toward real-time control applications.

% \section{Conclusions}
% \label{sec:Conclusions}
% % 本研究提出了一种基于ALE流固耦合计算框架的机器学习流固耦合计算方法，用于高效的流固耦合计算。本方法利用GNO高效处理非结构数据、ViT高效捕捉时序特征、边界修正方法和长序列训练方法引导网络学习，并在涡激振动柔性梁问题上，我们验证了该方法预测流固耦合的能力，结果表明，训练的模型能高效实现准确、长时序、稳定的流固耦合预测，而且进行的内插和外插泛化结果表明了此方法具有较好的泛化性。并进行了一系列的消融研究，主要结论如下：
% % 1.ViT模块能够高效捕捉时序特征。
% % 2.边界修正提升对流场固体边界物理量的预测效果。
% % 3.长序列训练。

% %
\section*{ACKNOWLEDGMENTS}
The authors gratefully acknowledge support from the National Key R\&D Program of China (2024YFC3013200).

\section*{AUTHOR DECLARATIONS}
The authors have no conflicts to disclose.

\section*{DATA AVAILABILITY}
The dataset of this study is available in the Zenodo repository~\cite{zhao2026_fsi_movies_zenodo}. 
Python code is available at \textcolor{blue}{\url{https://github.com/hunger-233/ale-gno-vit-fsi}}.

\appendix
\setcounter{figure}{0}
% \section*{Appendix}
\section {Additional GNO-ViT FSI prediction results}
\label{sec:appendix}
% x2 = -2 时，t1 u v p

\begin{figure}[!htbp]
    \centering
    \includegraphics[width=1\linewidth]{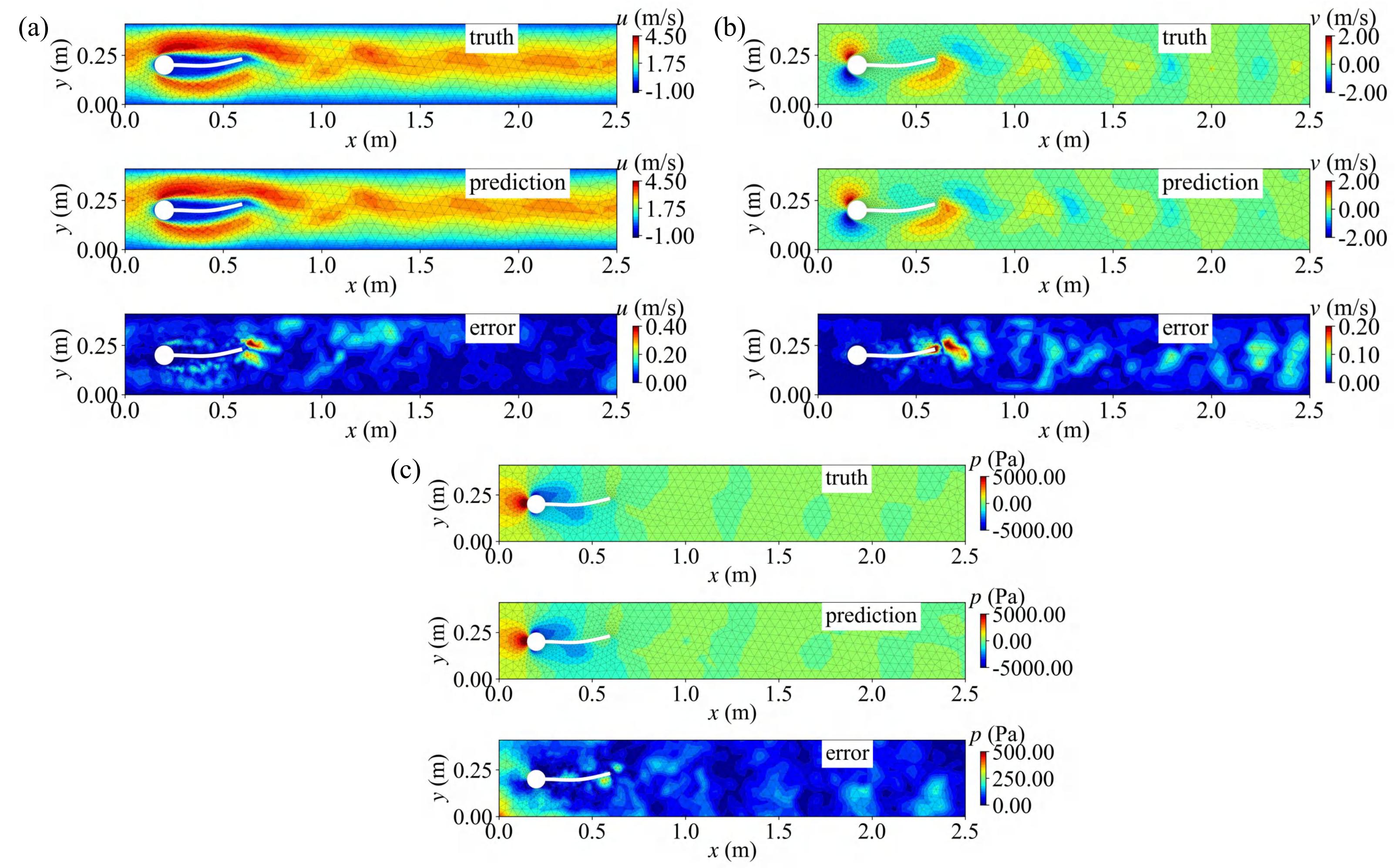}
    \caption{Comparison of prediction and truth of flow field at time $t_1$ for $c_2$=-2: (a) $u_f$, (b) $v_f$, (c) $p_f$.}
    \label{fig:contour-c-2-t1}
\end{figure}

\begin{figure}[!htbp]
    \centering
    \includegraphics[width=1\linewidth]{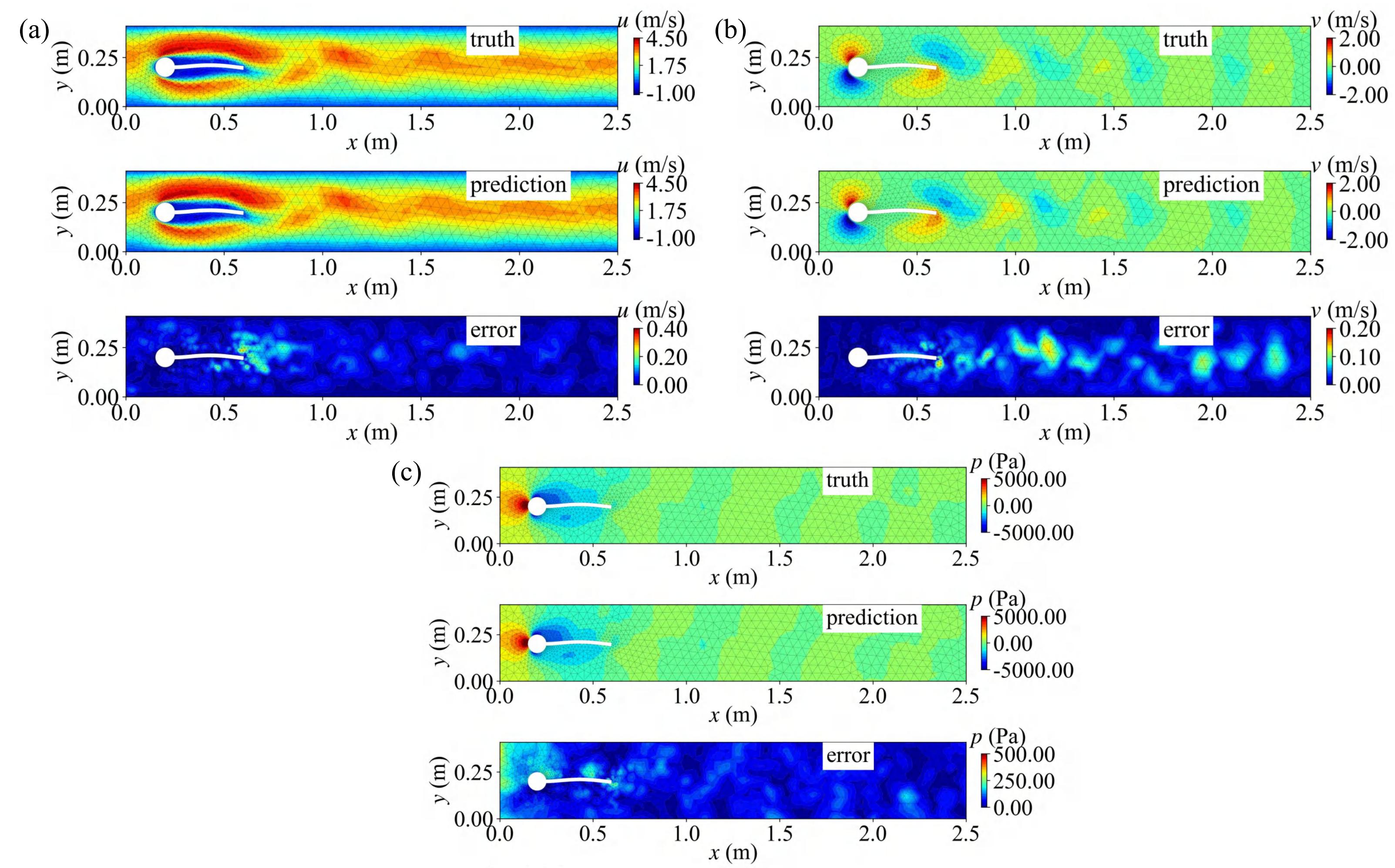}
    \caption{Comparison of prediction and truth of flow field at time $t_3$ for $c_2$=-2: (a) $u_f$, (b) $v_f$, (c) $p_f$.}
    \label{fig:contour-c-2-t3}
\end{figure}

\begin{figure}[!htbp]
    \centering
    \includegraphics[width=1\linewidth]{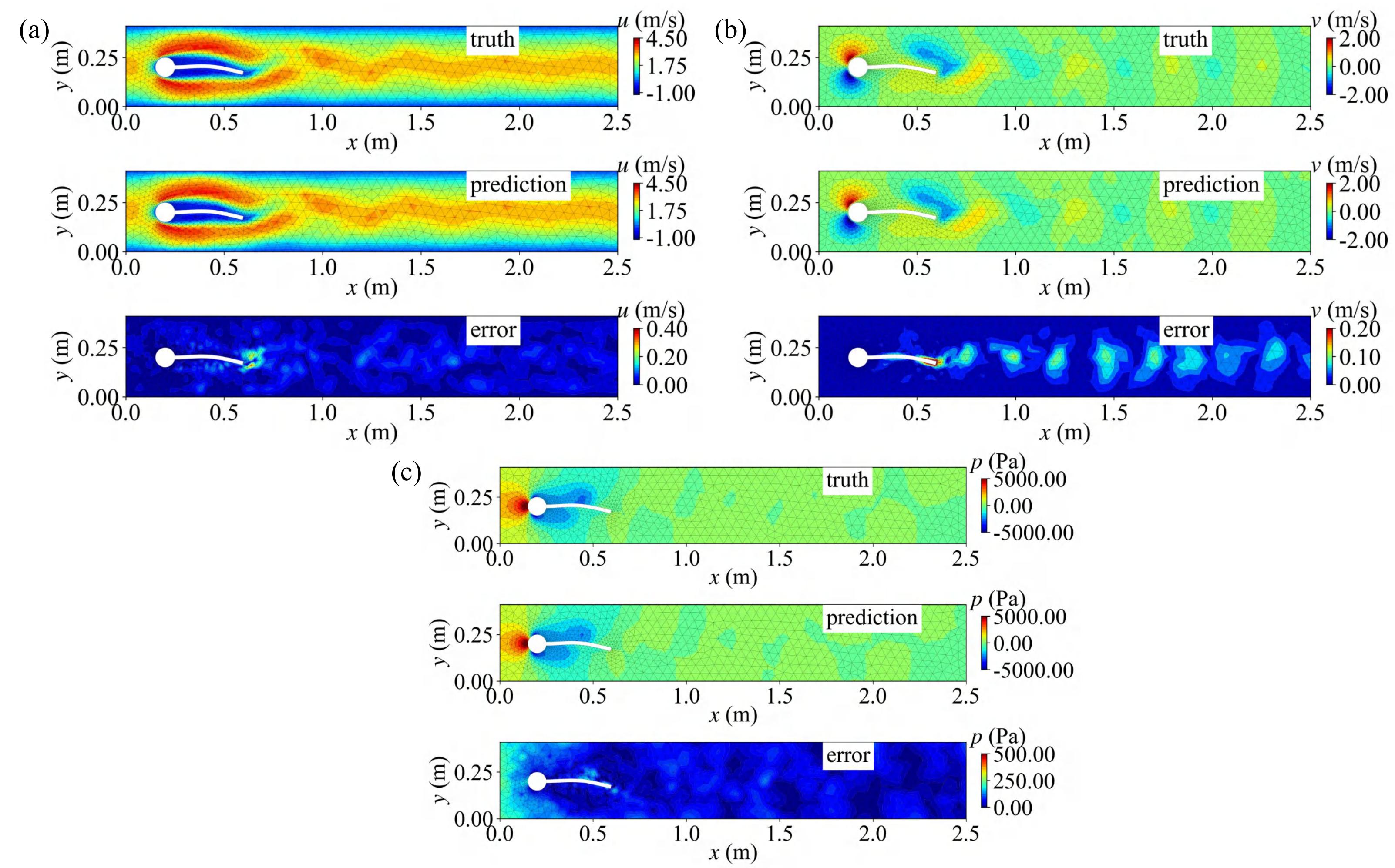}
    \caption{Comparison of prediction and truth of flow field at time $t_1$ for $c_2$=6: (a) $u_f$, (b) $v_f$, (c) $p_f$.}
    \label{fig:contour-c6-t1}
\end{figure}

\begin{figure}[!htbp]
    \centering
    \includegraphics[width=1\linewidth]{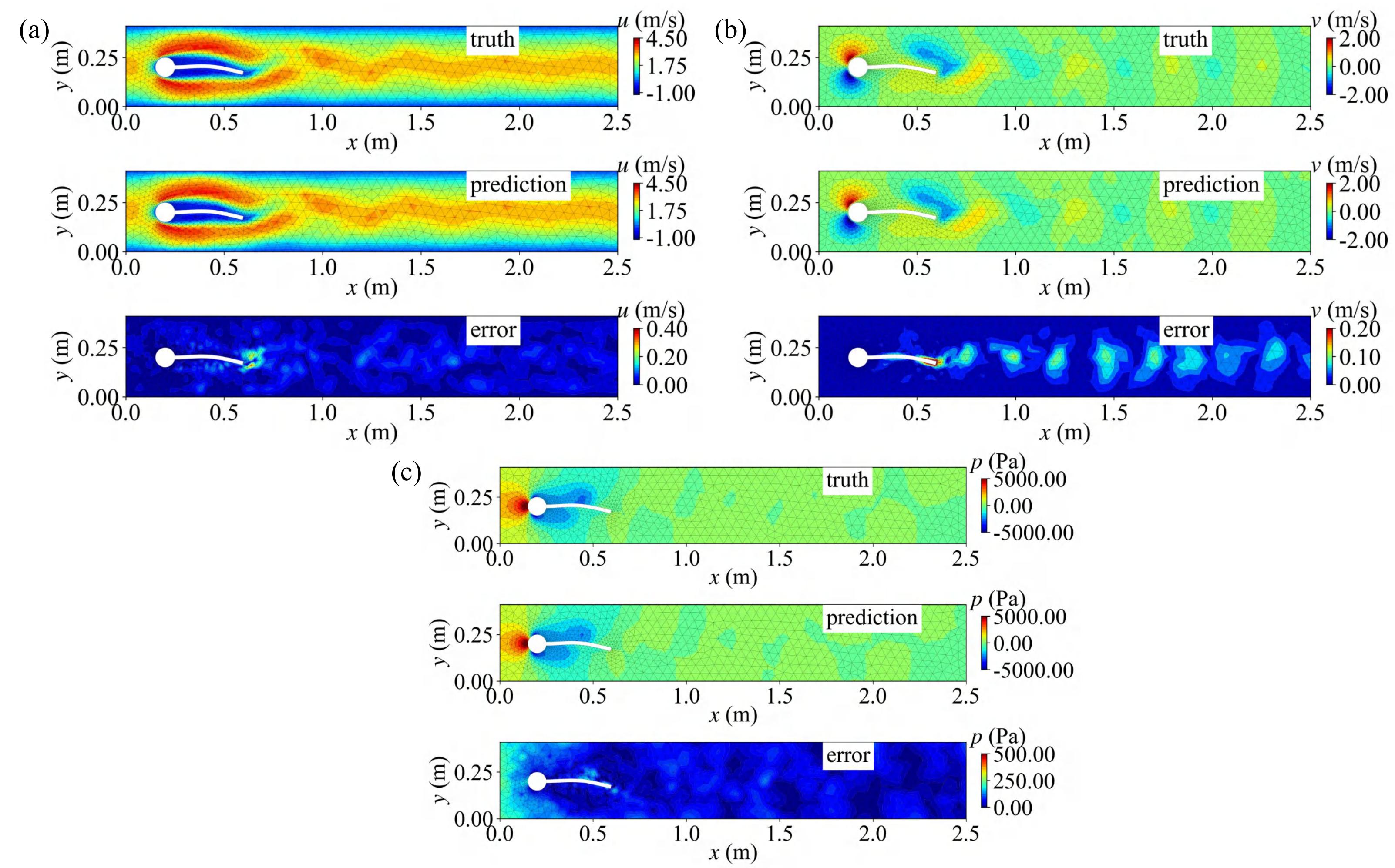}
    \caption{Comparison of prediction and truth of flow field at time $t_2$ for $c_2$=6: (a) $u_f$, (b) $v_f$, (c) $p_f$.}
    \label{fig:contour-c6-t2}
\end{figure}

\begin{figure}[!htbp]
    \centering
    \includegraphics[width=1\linewidth]{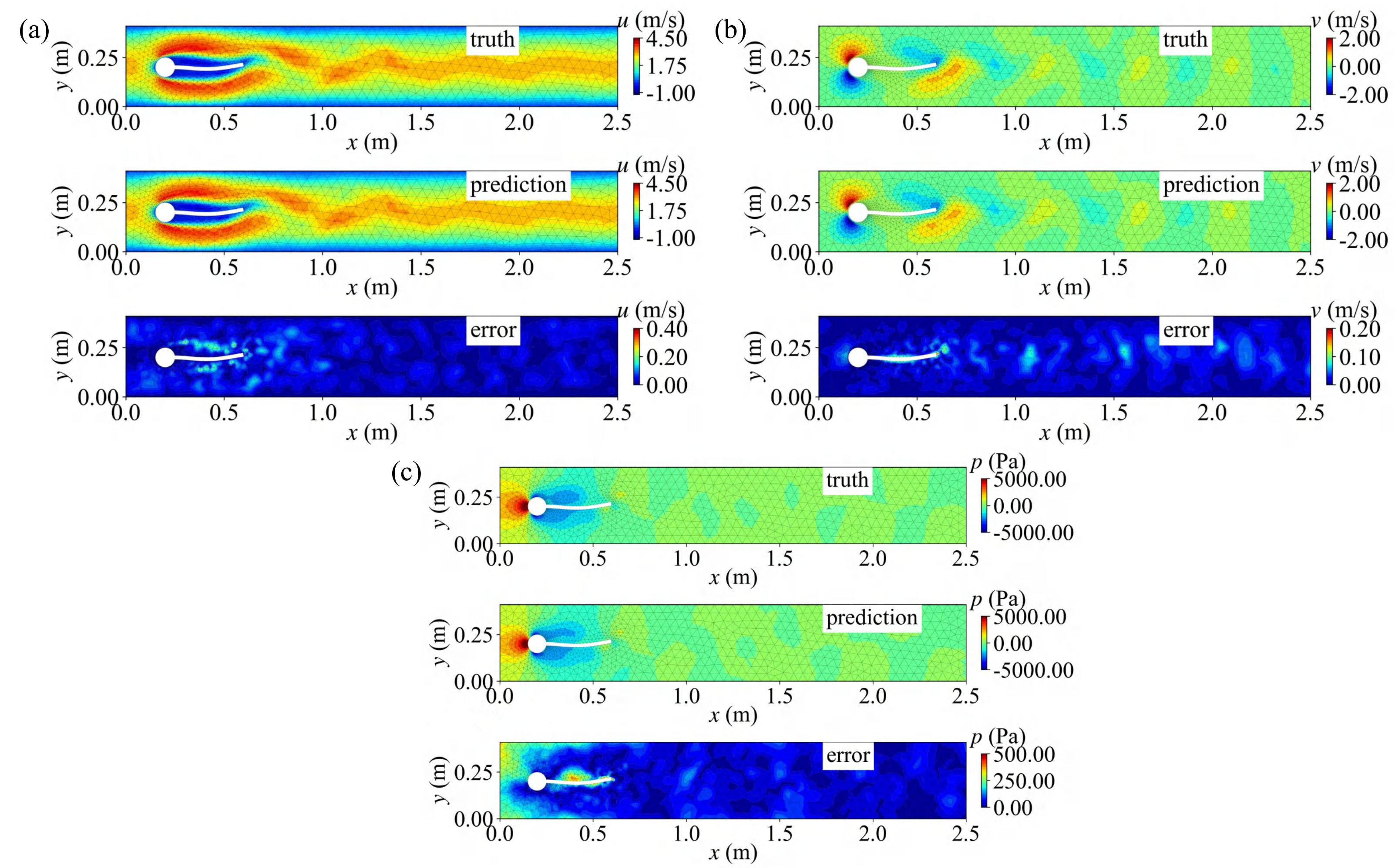}
    \caption{Comparison of prediction and truth of flow field at time $t_3$ for $c_2$=6: (a) $u_f$, (b) $v_f$, (c) $p_f$.}
    \label{fig:contour-c6-t3}
\end{figure}

\bibliographystyle{unsrtnat}
\bibliography{Reference}  % 文件名：Reference.bib

\end{document}